\documentclass[iop,apj]{emulateapj}
\usepackage{natbib}
\usepackage{graphicx}
\usepackage{booktabs}

\newcommand{\Frac}[2]{\frac{\displaystyle\strut #1}{\displaystyle\strut #2} }

\shorttitle{New associations of Fermi sources}
\shortauthors{Schinzel et al.}

\makeindex
\citeindextrue

\received{August 26, 2014}
\accepted{January 23, 2015}
\journalinfo{Astrophysical~journal Supplement Series}
\submitted{}

\begin{document}

\title{New Associations of Gamma-Ray Sources from the Fermi Second Source Catalog}

\author{Frank K. Schinzel\altaffilmark{1}, Leonid Petrov\altaffilmark{2}, Gregory B. Taylor\altaffilmark{1,3}, Elizabeth K. Mahony\altaffilmark{4}, \\ Philip G. Edwards\altaffilmark{5}, Yuri Y. Kovalev\altaffilmark{6,7}}

\altaffiltext{1}{Department of Physics and Astronomy, University of New Mexico, Albuquerque NM, 87131, USA}
\altaffiltext{2}{Astrogeo Center, Falls Church, VA 22043, USA}
\altaffiltext{3}{An Adjunct Astronomer at the National Radio Astronomy Observatory.}
\altaffiltext{4}{ASTRON, Netherlands Institute for Radio Astronomy, Postbus 2, NL-7990 Dwingeloo, the Netherlands}
\altaffiltext{5}{CSIRO Astronomy and Space Science, PO Box 76, Epping, NSW 1710, Australia}
\altaffiltext{6}{Astro Space Center of Lebedev Physical Institute, Profsoyuznaya 84/32, 117997 Moscow, Russia}
\altaffiltext{7}{Max-Planck-Institut f\"ur Radioastronomie, Auf Dem H\"ugel 69, D-53121 Bonn, Germany}

\email{Contact: fsch@unm.edu}

\begin{abstract}

We present the results of an all-sky radio survey between 5 and 9 GHz of 
sky areas surrounding all unassociated $\gamma$-ray objects listed in the 
\textit{Fermi} Large Area Telescope Second Source Catalog (2FGL). The 
goal of these observations is to find all new $\gamma$-ray AGN associations with
radio sources $>$10\,mJy at 8\,GHz. We observed with the Very Large Array and the Australia 
Telescope Compact Array the areas around unassociated sources, providing localizations of weak radio point sources 
found in 2FGL fields at arcmin scales. Then we followed-up a subset 
of those with the Very Long Baseline and the Long Baseline 
Arrays to confirm detections of radio emission on parsec-scales.
We quantified association 
probabilities based on known statistics of source counts and assuming a uniform 
distribution of background sources. In total we found 865 radio sources at arcsec 
scales as candidates for association and detected 95 of 170 selected for follow-up 
observations at milliarcsecond resolution. Based on this we 
obtained firm associations for 76 previously unknown $\gamma$-ray AGN. Comparison of these new 
AGN associations with the predictions from using the WISE color-color diagram 
shows that half of the associations are missed. We found that 129 out of 588 observed $\gamma$-ray sources 
at arcmin scales not a single radio continuum source was detected
above our sensitivity limit within the $3\sigma$ $\gamma$-ray localization. 
These ``empty'' fields were found to be particularly concentrated 
at low Galactic latitudes. The nature of these Galactic $\gamma$-ray 
emitters is not yet determined.

\end{abstract}


\section{Introduction}

The Large Area Telescope (LAT) aboard the \textit{Fermi} satellite has
been continuously observing the $\gamma$-ray sky since August 2008. So
far, this has resulted in the release of four all-sky catalogs, with a
fifth one in preparation. They all have in common a significant
fraction, typically $>30\%$, of the detected point sources with no
known counterpart at any other wavelength. To learn more about these objects we searched the radio sky
in the vicinity of \textit{every} unassociated $\gamma$-ray source
deeper than any survey has done before to find new counterparts at other 
wavelengths in order to understand the nature of those currently unassociated 
objects.

The previous $\gamma$-ray experiment in orbit covered an energy
range of 20 MeV -- 30 GeV, slightly less in energy than that of
\textit{Fermi}/LAT (20 MeV -- $>$300 GeV). This was the Energetic Gamma-Ray
Experiment Telescope (EGRET) on board the Compton-Gamma Ray Observatory
(CGRO). Among the 188 point sources found by EGRET, there were 87 (46\%) for
which no multi-wavelength counterpart was found by cross-matching
catalogs with expected $\gamma$-ray emitting sources such as pulsars,
blazars, or supernova remnants \citep{2008AA...489..849C} and using
other parameters such as correlated variability. This was complicated
by the poor angular resolution (5--$30'$) and limited field-of-view
($\sim20^\circ$) of EGRET.

\begin{figure}[htbp!]
  \centering
  \includegraphics[angle=-90,width=\columnwidth]{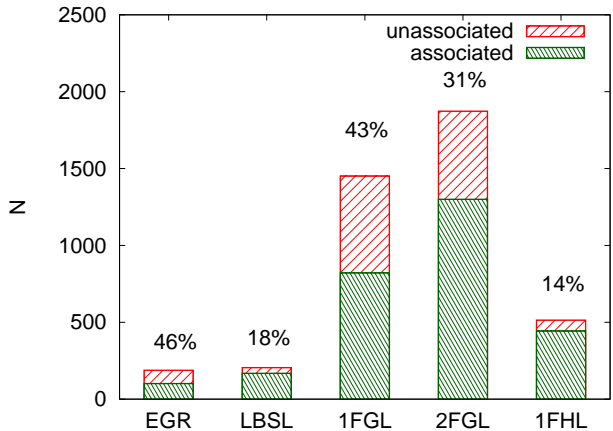}
  \caption{Comparison of associated and unassociated source counts of all
           major $\gamma$-ray source catalogs since EGRET.\label{fig:1}}
\end{figure}

The latest generation of $\gamma$-ray space telescope,
\textit{Fermi}/LAT, provided a significant advancement for studying the
$\gamma$-ray sky. However, despite recent efforts to identify those
unassociated $\gamma$-ray emitters with known sources, the number of
unassociated objects remains high throughout the \textit{Fermi}-era.
Fig.~\ref{fig:1} illustrates the fraction of unassociated $\gamma$-ray
sources for all the all-sky catalogs released so far, the revised
EGRET point source catalog \citep[EGR;][]{2008AA...489..849C} and the
\textit{Fermi} 3-month bright source list
\citep[LBSL;][]{2009ApJS..183...46A}, 11-month point-source catalog
\citep[1FGL;][]{2010ApJS..188..405A}, 2-year point-source catalog
\citep[2FGL;][]{2012ApJS..199...31N}, and first hard source list
\citep[1FHL;][]{2013ApJS..209...34A}. For the major catalogs, 1FGL and
2FGL, the fraction of unassociated sources has never dropped below
30\%. Sources listed in LBSL and 1FHL are largely associated with bright radio-loud
AGN for which catalogs are mostly complete thus the fraction of unassociated
sources ($<20\%$) is partly due to this selection effect.

The 1-$\sigma$ positional uncertainty of the 2FGL catalog ranges
from $8''$ to $15'$ with a median
value of $2.5'$. Such large position errors makes close to impossible
the direct association of $\gamma$-ray sources with optical data
without considering other information (e.g. variability), since many
candidates for association can be found within the $\gamma$-ray
source localization errors. From the analysis of early EGRET data it
was found that $\gamma$-ray emission and parsec-scale radio emission
are strongly related, e.g. \citep{1998ApJ...500..673T}. An early comparison
of \textit{Fermi} data with radio sources from Very Long Baseline Interferometry (VLBI)
observations using the 3-month list \citep{2009ApJS..183...46A} has shown that parsec-scale emission detected
through VLBI is one of the best and most efficient tools to both associate \textit{Fermi}
AGN and to understand the probability of false detections \citep{2009ApJ...707L..56K}.
Cross-correlation of 2FGL against a cumulative list of radio
sources\footnote{Available at \url{http://astrogeo.org/rfc}} detected
in dedicated VLBI surveys at the Very Long Baseline Array (VLBA),
Australian Long Baseline Array (LBA), and European VLBI Network (EVN)
\citep{2002ApJS..141...13B,2003AJ....126.2562F,2005AJ....129.1163P,2006AJ....131.1872P,2007AJ....133.1236K,
  2008AJ....136..580P,2011MNRAS.414.2528P,2011AJ....142...35P,2012MNRAS.419.1097P,2007ApJ...671.1355T,2011AJ....142...89P,
  2009JGeod..83..859P,2011AJ....142..105P,2013AJ....146....5P,2011arXiv1110.6252C}
revealed that for \textit{the majority} of 2FGL sources, 1081 out of
1872 objects have a radio counterpart with emission at milliarcsec
scales brighter than 10\,mJy at 8\,GHz. All but three of these sources
(Vela pulsar and magnetars/microquasars LSI61303 and LS5039) are active 
galactic nuclei (AGN). Therefore, an association of a $\gamma$-ray source 
with a compact radio source statistically implies its classification as an AGN.

Had a VLBI cumulative catalog been complete to a 10\,mJy level, we
would have been able to establish or rule out radio-$\gamma$
association of AGN for almost every 2FGL source. However by 2014,
completeness of the all-sky for 8~GHz VLBI catalogs at the 10\,mJy
level has reached only 9\%. Realizing the importance of radio observations for
$\gamma$-ray associations, we launched several all-sky campaigns for
observing areas within error ellipses of $\gamma$-ray sources without
an association. The goal of these campaigns is to find all $\gamma$-ray
AGN associations with radio sources brighter than 10~mJy at
8~GHz. This will allow us to study two interesting issues: firstly, a detailed
analysis of the population of $\gamma$-ray AGN; and secondly, the
population of remaining unassociated sources that show no radio
emission.

In the following we present the results from our observing program
searching for faint radio counterparts among all unassociated
$\gamma$-ray sources listed in 2FGL. Section~\ref{sec:2} describes the
observing program and data analysis procedures for each observatory
from which data were acquired.  This is followed by Section~\ref{sec:3}
presenting results and the full catalog of detected radio sources together
with new associations of $\gamma$-ray sources. This is followed by a discussion 
of our findings in Section~\ref{sec:4} including a comparison with alternative 
association methods. A brief conclusion and summary is provided in
Section~\ref{sec:5}.  In the following, the radio spectral
index $\alpha$ is defined as $S \propto \nu^\alpha$, where $S$ is the 
source flux density and $\nu$ is the observing frequency.

\begin{center}
\begin{table*}[!hbtp]
  \centering
    \caption{List of observations from which results are presented.\label{tab:observations}}
    \small
  \begin{tabular}{lllcrccrl}
    \hline
     Telescope & Config. & Code & Start & Dur.     & Tune & B/w    & \#  & Identifier\\
       &   &      & (UTC) & (h)      & (GHz)& (GHz)  &   \\
    \hline
   ATCA      &  H214   & C2624 & 2012 Sep 19 10:00 & 29 & 5.5/9.0 & 2.0           & 411  & AOFUS1\\       
   VLA       &    A    & S5272 & 2012 Oct 26 11:19 & 2  & 5.0/7.3 & 1.0           & 41   & VOFUS1A\\       
   VLA       &    A    & S5272 & 2012 Nov 03 21:01 & 7  & 5.0/7.3 & 1.0           & 175  & VOFUS1B\\
   VLBA      &    -    & BP171 & \multicolumn{2}{c}{2013 Feb -- 2013 Aug} & 4.1/7.4 & 0.48         & 56   & VCS7\\
   LBA       &    -    & V271  & \multicolumn{2}{c}{2013 Mar -- 2013 Jun} & 8.2     & 0.13         & 21   & LCS2\\ 
   VLBA      &    -    & S5272A & 2013 Jun 08 06:12 & 12 & 7.4 & 0.48           & 33   & VOFUS2A\\
   ATCA      &  H214   & C2624  & 2013 Sep 25 21:30 & 45 & 5.5/9.0 & 2          & 997  & AOFUS2\\
   VLBA      &    -    & S5272B & 2013 Oct 19 05:41 & 12 & 7.4 & 0.48           & 20   & VOFUS2B\\
   VLBA      &    -    & S5272C & 2013 Dec 02 16:55 & 12 & 7.4 & 0.48           & 33   & VOFUS2C\\
   VLBA      &    -    & S5272D & 2013 Dec 05 13:49 & 12 & 7.4 & 0.48           & 31   & VOFUS2D\\
    \hline
   
  \end{tabular}
  \begin{flushleft}\footnotesize
    Column descriptors: 
    Config. -- array configuration; Code -- observation proposal code;
    Start -- start time in UTC; Dur. -- duration of the observation in hours;
    Tune -- center frequency of the tunings; B/w -- bandwidth of each tuning;
    \# -- number of targets observed, in the case of AOFUS2 the number of 
          pointings performed in mosaicking mode; Identifier -- custom observation identifier.\\[2pt]
     Note: VCS7 and LCS2 campaigns had many segments. Target sources were 
            piggy-backed onto these campaigns, so only a date range for these
            observations is given instead of the observation start time.\\
  \end{flushleft}

\end{table*}
\end{center}

\section{Observing Program}\label{sec:2}

The observing program was executed in three steps. In the first step
we observed with the Australia Telescope Compact Array (ATCA) and the
Karl G. Jansky Very Large Array (VLA) every unassociated source using
the 5--10 GHz frequency range. The telescopes were pointed to 2FGL
positions. The field-of-view was defined by the area within which the total power
does not drop below 20\% with respect to the pointing direction. The
radius from the pointing center at which a 20\% drop occurs for the ATCA
22\,m diameter antennas is $6.5'$ and $4.0'$ at 5.5 and 9.0\,GHz respectively
and for the 25\,m diameter VLA antennas is $6.2'$ and $4.3'$ at 5.0 and 7.3\,GHz
respectively. Thus the field of views can be considered comparable. This 
allows imaging of the entire \textit{Fermi}/LAT localization error ellipses, 
which typically have a 1-$\sigma$ semi-major axis of 2--$4'$. The 
angular resolution of our ATCA and VLA observations differs significantly: 
the primary synthesized beam is $0.3''$ at the VLA and $40''$ at the ATCA.

The second step involved only ATCA observations, where we re-observed
the fields A)~where sources were detected beyond the $6.5'$ field of
view or within a sidelobe; B)~where no source has been detected. In
case A we pointed the array to the direction of the previous
detection, in case B we used a hexagonal mosaic to double the field of
view.

In a third step, we followed-up with VLBI, those sources detected with ATCA
and VLA that are brighter than 10~mJy at any sub-band within
5--9~GHz. In total, 339 sources fit this criteria out of 865 detected
objects in VLA and ATCA observations. We applied a flux density cutoff 
of 10~mJy to increase the success for a detection given limited observational
resources. In addition, the probability to find a background compact
radio source weaker than 10\,mJy within the \textit{Fermi} error 
ellipse becomes non-negligible. 

We used the Australian Long Baseline Array (LBA) with baselines up to
1700~km long and the Very Long Baseline Array (VLBA) with baseline
lengths up to 8,600~km. Although it is technically feasible to
observe the \textit{Fermi} fields with VLBA/LBA directly in one step,
we adopted a multi-step approach for the optimal use of the resources
available. We added 21 relatively bright target sources to the
on-going programs, the LBA Calibrator Survey \citet{2011MNRAS.414.2528P} in
the southern hemisphere, and 46 sources to the VLBA Calibrator
Survey--7 in the northern hemisphere. These 67 objects fit the
goals of these surveys. In addition we ran a dedicated VLBA observing
program (VOFUS2) for observing the remaining targets with declinations $>
-30^\circ$. A summary of all observations and their characteristic parameters is presented
in Table~\ref{tab:observations}. 

\pagebreak

\subsection{Very Large Array}

A list of 216 target fields were observed with the VLA using a total
bandwidth of 2 GHz, recording both left and right handed circular
polarizations, with the integration time set to 1\,s. Real-time
correlation was performed using VLA's WIDAR (Wideband Interferometric
Digital ARchitecture) correlator. The instantaneous bandwidth was
split into two parts, with one half centered at 5.0 GHz
(4.5--5.5\,GHz) and the other centered at 7.3 GHz (6.8-7.8\,GHz). This
provides simultaneous observation of two separate frequency bands.
The observing time of 9 hours was split into two segments to be able
to observe fields at all LST ranges (see
Table~\ref{tab:observations}).  The first segment started on 2012
October 26, using 26 antennas. One antenna did not provide fringes in
two of its intermediate frequencies (IFs) for the duration of the
observing segment, affecting the sub-band range of 6.8--7.8 GHz
only. The second segment started on 2012 November 3 and recorded with
27 antennas of the array. One of the antennas recorded with zero
fringe amplitudes in two of the four IFs, providing for this antenna
only half of the available bandwidth in each of the sub-bands for the
entire duration of this segment. All array data were lost during
correlation for the scan centered on the position of
2FGL\,J0423.4+5612.

At the beginning and end of each observing segment a bright flux
density/bandpass calibrator was observed.  In the case of the first
segment these were 3C\,147 and 3C\,286 and, in segment two 3C\,286 and
3C\,48. Each target source was observed only once with a total
integration time of $\sim$45\,s. Nearby phase calibrators were added
with typical integration times of 30\,s each in order to be able to
solve for changes in the complex gains during the target observations.
For both segments the VLA was in A array configuration, providing
baseline lengths from 0.68 to 36.4\,km, which results in sub-arcsecond
resolutions and a field of view of up to about 7 arcmin.

Initial calibration was performed using the Common Astronomy Software
Applications (CASA) release 4.0.0\footnote{\url{http://casa.nrao.edu}} and 
the Astronomical Image Processing System\footnote{\url{http://www.aips.nrao.edu/index.shtml}} 
\citep[AIPS;][]{1990apaa.conf..125G}. Before any calibration was performed the
observations were inspected for radio frequency interference (RFI) that
degrade the data. Only minimal data flagging was required for the frequency
range 4.5--5.5\,GHz, whereas the frequency range 6.8--7.8\,GHz was
significantly affected by strong RFI. In order to cope with Gibbs
ringing introduced by strong RFI, the data covering 6.8--7.8\,GHz were
smoothed using the Hann window function implemented in the CASA task
\texttt{hanningsmooth}\footnote{This function uses a Hann window to smooth
the frequency channels with a weighted running average, which supresses the
oscillations introduced by the Fourier transform of a strong narrowband RFI signal.}. 

After the initial flagging of RFI affected channels and times,
corrections for system temperature variations measured by noise diodes
installed at VLA receivers were determined and applied to the data
using the tasks \texttt{gencal} (with \texttt{caltype='evlagain'}) and
\texttt{applycal}. The corrected data column was then written out with
the task \texttt{exportuvfits} to be imported into AIPS which provides
the possibility to perform a global fringe fit, which is not
yet implemented in CASA. Within AIPS the global fringe fit (task
\texttt{FRING}) using the primary flux calibrator (3C\,147 or
3C\,286) provides correction factors to flatten the phases across the
passband which are applied to the entire dataset. After applying the
corrections to the dataset using AIPS tasks \texttt{CLCAL} and
\texttt{SPLAT}, the data were exported from AIPS and imported back into
CASA for the final calibration and imaging steps.

In this last step the VLA antenna elevation dependent gain and antenna
efficiencies were corrected for. The flux densities for the primary
flux calibrator were determined using the corresponding models for
3C\,147 and 3C\,286 which are provided in CASA\footnote{Also available
  at:
  \url{https://science.nrao.edu/facilities/vla/data-processing/flux-calibrator-models-for-new-evla-bands}}.
After this the bandpass calibration was performed using the primary and
secondary flux calibrators (3C\,147, 3C\,286, and 3C\,48).  Finally
complex gain corrections were derived for the secondary calibrators and
the flux density scale was determined.  In the last step all these gain
and phase corrections were applied to the entire dataset including the
target fields.  The resulting dataset was inspected and channels around
the bandpass edges were flagged.

All targeted source fields were deconvolved using the CASA task
\texttt{clean}. This task uses a Clark based clean algorithm
\citep{1980AA....89..377C}, applies a $w$-projection needed for
widefield imaging using 100 projection planes, and uses visibility weights
determined according to Briggs weighting scheme \citep{PhDT...1995B}.
The deconvolution was run with 5000 iterations, a default
loop gain of 0.1, and a flux density threshold at which to stop
cleaning of 0.05\,mJy corresponding to the approximate thermal noise
limit. All target fields were imaged assuming a flat spectral index
across the 1\,GHz bandwidth. All images are then analyzed by a
semi-automatic script reading images into AIPS and using the task
\texttt{SAD} to identify point sources.  Custom scripts in the Python
programming language were developed to analyze the resulting images and point sources
found. Sometimes the algorithm applied in \texttt{SAD} identifies
image artifacts in the case of strong point-sources, which were
manually flagged in order to retain a clean list of point sources that
includes position and flux density for each of the two 1\,GHz
sub-bands which are used in subsequent analysis.

\subsection{Australia Telescope Compact Array}

During the first campaign with ATCA, from 2012 September 19--20,
we observed 411 2FGL unassociated sources in a declination range of 
$[-90\degr, +10\degr]$ at 5.5 and 9~GHz for 29 hours. The details of that
observing campaign and results have been reported by
\citet{2013MNRAS.432.1294P}. We detected a total of 424 point sources.

In a second 45~hour ATCA campaign on 2013 September 25--28 we
re-observed sources that were detected at 5 GHz, but were not detected
at 9~GHz, and had position offsets exceeding the full-width at half
maximum of the 9~GHz beam in the first ATCA campaign. This included
also sources that were detected at the edge of the beam or within a
sidelobe. In addition, 130 sources that were not detected in
2012 were re-observed using the six-element mosaic mode allowing detection 
of all radio sources brighter than 1\,mJy located within the entire 95\%
probability ellipse of the 2FGL localization.

Both the 2012 and 2013 observations were conducted in array
configuration H214 with baselines ranging from 82--247\,m between the
inner five antennas and $\sim$4.4\,km between the sixth, CA06, and the inner
antennas. Observations were recorded simultaneously in two bands
centered at 5.5 and 9.0\,GHz with a bandwidth of 2\,GHz each and in
both linear polarizations.

\subsection{Very Long Baseline Array}

Follow-up observations of 149 targets selected from the VLA and ATCA
survey above -30$^\circ$ declination were conducted with the
VLBA. Sources with VLA flux densities greater than 20~mJy were added to the
list of targets of the VLBA Wide-Field Calibrator Survey-7 (VCS7,
L. Petrov, paper in preparation).  These observations were scheduled
in the so-called filler mode by the array operator using software with
a web front-interface that we provided.  The array operator scheduled
segments with blocks 3--8 hours long during periods of time when one
or two stations were down due to technical problems or suffered from
bad weather that impeded high-priority projects. The goal of the VCS7
project was to find more compact radio sources, to determine their
positions with sub-mas accuracies, and to obtain snap-shot images.
Observations of strong pre-selected targets fit the goal of that
campaign.  VCS7 observations were taken in right circular polarization
at two wings of the C-band receiver: at 4.128--4.608 and
7.392--7.872~GHz simultaneously.  Each subband was recorded in 8 IFs
of 32~MHz each with some gaps. The aggregate recording rate was
2~Gbps. Among 94 candidates inserted in the observation queue, 48 were
observed in one scan 180~s long each, and 29 were detected. 
Sources that have not been detected have a correlated flux density weaker 
than 12~mJy.

The remaining 101 sources with declinations $>-30^\circ$ and flux densities 
$>10$\,mJy were observed in a dedicated campaign S5272
in 2013 June--December in four 12 hour long segments (see
Table~\ref{tab:observations}). These observations used only the upper 
sub-band (7.392--7.872~GHz) with the same intermediate frequency setup as in
VCS7, but ran in the dual polarization mode. This lowered the detection 
limit by 40\%. Each target was observed in 3--4 scans of 210 seconds long
each, along with a suitable phase calibrator located within 2--$3\degr$.

The data were correlated with the NRAO's DiFX software correlator in
Socorro, NM \citep{2011PASP..123..275D}. The spectral resolution of
the output visibility function, 0.25~MHz, and duration of accumulation
periods, 0.25~s, was sufficient to detect sources within $1'$ of the
field center.  Fringe fitting over the entire search window was
performed with \textit{PIMA} software\footnote{See
  \url{http://astrogeo.org/pima}}. The result of fringe fitting, group
delays and the visibilities averaged over frequency and over 20~s long
time intervals were exported for post-processing with
VTD/Post-Solve\footnote{See \url{http://astrogeo.org/vtd}} and Difmap data processing software for absolute 
astrometry analysis and imaging respectively. Astrometric analysis consists of 
two steps: initial and final. During the initial step a parametric model that includes
coordinates of target sources, positions of stations, and the clock function
in residual path delay represented as a B-spline of the 1st degree,  was
adjusted to group delays using least squares fitting. At first, only
the observations with the probability of a false detection of less
than 0.01 were used. For this experiment observations with the 
signal-to-noise ratio (defined as the ratio of the fringe amplitude to the
average value of noise) greater than 5.76 satisfied that criteria. We
ran an iterative procedure for outlier elimination. Observations with
residual group delay exceeding $4\sigma$ were discarded. Sources with
less than four remaining observations were considered
non-detections. We have detected a little over one half 
of the sources from our S5272 
campaign: 58 out of 101. The procedure of fringe fitting with a narrow search
window, 2~ns over group delay and $10^{-12}$ over delay rate, was
repeated for detected sources.  Then the step of initial astrometric
analysis was repeated. The final astrometric analysis was done by
running a global least square solution with all astrometry/geodesy
observations acquired since 1980 April 1 through 2014 April 7 including
VCS7 and S5272 experiments, a total of 10.7 million observations. In the case of
astrometric/geodetic analysis an observation refers to a single group delay obtained
from one scan (pointing) of one baseline. Then
the observations of detected sources, except those which were marked
as outliers during an astrometric solution, were averaged over
frequencies within an IF and over 20 seconds and used for hybrid
imaging with Difmap. The post-processing procedure is described in more
detail in \citet{2011AJ....142...35P}.

Astrometric accuracy for all but two sources was in the range of 0.2--2 mas,
with a median uncertainty of 0.4~mas. The image rms was in the range 70--120
$\mu$Jy and it was limited by thermal noise. The overall baseline-based detection 
limit was around 8~mJy. 

\subsection{Long Baseline Array}

For sources with declinations below -$30\degr$ we added 21 objects to
the on-going LBA Calibrator Survey (LCS) campaign
\citep{2011MNRAS.414.2528P}.  In the LCS campaign we observed at
8.200--8.520~GHz, but using different stations equipped with
heterogeneous data acquisition systems recording at different bandwidth
due to hardware limitations. The recording rate ranged from 256 to
1024~Mbps for different stations. Target sources were observed in two
scans of 240~s long each. The astrometric analysis performed was similar to
that used in the VCS7 and S5272 campaigns, except that it involved an
additional step of resolving group delay ambiguity with a spacing of
3.9~ns using a highly sophisticated algorithm. Since it is problematic
to produce high quality snap-shot images using the LBA network, we
resorted to evaluation of median correlated flux densities in three
ranges of projected baseline lengths: 0--6~M$\lambda$,
6--25~M$\lambda$, 25--50~M$\lambda$. The details of analysis of LBA
observations can be found in \citet{2011MNRAS.414.2528P}.

Of 21 sources observed on 2013 March 15 and 2013 June 15, we have
detected 8 objects. The detection limit was $\sim$10~mJy.  Thus, in
three VLBI campaigns we have to date observed half of the candidates 
for association that were found and detected 95 sources, i.e., 56\%. We plan 
to observe the remaining 159~objects in the near future.

\section{Results}\label{sec:3}

\subsection{Radio Source Catalog}

For ATCA the root-mean-squared (rms) of image noise was found to be in
the range of 0.15--0.25\,mJy beam$^{-1}$, with a typical full width at
half-maximum (FWHM) size of the restored beam of $35''$ at 5\,GHz
and $20''$ at 9\,GHz. The detection limit was determined to be
1.8\,mJy for sources in the center of the field of view and 9\,mJy at
the edge of the 5.5\,GHz field of view, at $6.5'$.  In the case
of the VLA observations, the image noise rms was found to be in the
range of 0.18--0.01\,mJy beam$^{-1}$ with a median value of 0.08\,mJy beam$^{-1}$, a
factor of two smaller than that of ATCA. The typical FWHM size of the
restored beam was $24''$ at 5.0\,GHz and $7''$ at 7.3\,GHz
providing a significantly higher resolution and a higher rejection of
extended sources than ATCA. The 2$\sigma$ detection limit for the VLA
observations was determined to be 0.37\,mJy for sources in
the center of the field of view and 1.1\,mJy at the edge of the 5.0
GHz field of view, at $6.5'$. Sources with evidence of extended
emission were flagged accordingly. The radio sources found in the ATCA
observation at 5.5 and 9.0\,GHz around 2FGL\,J1634.4-4743c were not
included in the following analysis due to indications of the presence
of an extended source structure and confusion from the Galactic plane.

The detected point source candidates for both the VLA and ATCA were
combined into a single uniform dataset and the positions determined at
different frequencies were spatially cross-matched in order to
identify detections of the same object across the observed bands. For
the 35 fields that were observed both with ATCA and the VLA, ATCA
information was only used if no point source was detected by the VLA
within $20''$ of the $\gamma$-ray localization. ATCA
observations are sensitive to larger angular scale structures as compared
to the VLA and have thus a higher chance of detecting extended Galactic 
sources. These fields were also used
to check for systematic differences between ATCA and VLA
observations. We found good agreement between the detected point
sources for most of those fields. However, for some sources the VLA
counterpart showed fainter emission indicating a significant component
of extended emission. There were also a small number of weak point
sources only detected by either ATCA or VLA but not by both. Since
ATCA is more sensitive to larger scale emission, due to shorter
baselines, those sources added by ATCA are most likely related to
large scale radio sources rather than to compact sources which are
primarily targeted. Nevertheless, we have included those detections
in the combined candidate catalog. To summarize, we have detected a total 
of 1,268 unique point sources with connected radio interferometers in 588
fields selected from the 2FGL catalog. Among those were 325 observed
with the VLA and 943 observed with ATCA, where 58 sources were found
in fields co-observed by the VLA and ATCA.

All radio point sources found were cross-matched against the NRAO VLA Sky
Survey \citep[NVSS:][]{1998AJ....115.1693C}, Sydney University
Molonglo Sky Survey \citep[SUMSS, version 2.1 of 2012 February
  16:][]{1999AJ....117.1578B,2003MNRAS.342.1117M}, the Molonglo
Galactic Plane Survey 2nd Epoch \citep[MGPS-2:
][]{2007MNRAS.382..382M}, the Supercosmos database \citep{2001MNRAS.326.1279H}, and the 
Wide-field Infrared Survey Explorer (WISE)
catalog\footnote{\url{http://wise2.ipac.caltech.edu/docs/release/allwise/}}
\citep[ALLWISE, November 13, 2013][]{2010AJ....140.1868W,
  2011ApJ...731...53M}, which combines the data from the WISE
cryogenic and post-cryogenic survey phases providing the most
comprehensive view of the full mid-infrared sky currently available.

  \begin{table*}[bt!]
  \centering
  \caption{The first 8 rows of 148 objects that were detected at 5.0/5.5 
           and/or 7.3/9.0 GHz within 2.7 arcmin of the 2FGL counterpart localization.
           Table~\ref{tab:catI} is published in its entirety in the electronic
           edition. A portion is shown here for guidance regarding its form and
           content.\label{tab:catI}}
  \tiny\begin{tabular}{@{}c@{\,\,}c@{\,\,\,}c@{\,\,\,}c@{\,\,\,}c@{\,\,\,}c@{\,}c@{\,\,\,}r@{\,}r@{\,}r@{\,\,\,}c@{\,}c@{\,}c@{\,\,\,}c@{\,\,}c@{\,\,\,}c@{\,}c@{\,\,\,}c@{\,}c@{\,\,\,}c@{\,}c@{\,\,\,}c}
\hline
IAU name & 2FGL name & F & $\alpha$ & $\delta$ & $\Delta\alpha$ & $\Delta\delta$ & $F_5$ & $F_7 $ & $F_9$ & $\Delta F_5$ & $\Delta F_7$ & $\Delta F_9$ & $Sp_5$ & $Sp_9$ & $\Delta Sp_5$ & $\Delta Sp_9$ & $Sp$ & $\Delta Sp$ & $D$ & $N\sigma$ & Camp. \\
         &      &      & (h) (min) (s) & ($\circ$) (') ('') & ('') & ('') & (mJy) & (mJy) & (mJy) & (mJy) & (mJy) & (mJy) & & & & & & & (') & & \\ 
(1) & (2) & (3) & (4) & (5) & (6) & (7) & (8) & (9) & (10) & (11) & (12) & (13) & (14) & (15) & (16) & (17) & (18) & (19) & (20) & (21) & (22) \\ 
\hline
J0002$+$6219 & J0002.7$+$6220 &  & 00 02 53.52 & $+$62 19 17.03 & 0.050 & 0.050 & 3.47 & 1.78 &  & 0.19 & 0.20 &  &  &  &  &  & -1.77 & 0.33 & 1.6 & 0.8 & V1\\
J0004$+$2206 & J0004.2$+$2208 &  & 00 04 07.36 & $+$22 06 15.76 & 0.050 & 0.050 & 1.86 & 1.38 &  & 0.18 & 0.20 &  &  &  &  &  & -0.79 & 0.46 & 2.4 & 0.6 & V1\\
J0039$+$4330 & J0039.1$+$4331 &  & 00 39 01.86 & $+$43 30 29.35 & 0.050 & 0.050 & 1.70 & 1.14 &  & 0.20 & 0.16 &  &  &  &  &  & -1.06 & 0.48 & 1.5 & 0.6 & V1\\
J0039$+$433A & J0039.1$+$4331 & f & 00 39 08.16 & $+$43 30 14.63 & 0.050 & 0.050 & 6.46 & 5.99 &  & 0.20 & 0.20 &  &  &  &  &  & -0.20 & 0.12 & 1.4 & 0.5 & V1\\
J0102$+$0944 & J0102.2$+$0943 & f & 01 02 17.11 & $+$09 44 09.54 & 0.050 & 0.050 & 14.77 & 14.72 &  & 0.18 & 0.17 &  &  &  &  &  & -0.01 & 0.04 & 1.2 & 0.4 & V1\\
J0103$+$1323 & J0103.8$+$1324 & f & 01 03 45.74 & $+$13 23 45.25 & 0.050 & 0.050 & 19.79 & 17.62 &  & 0.15 & 0.15 &  &  &  &  &  & -0.31 & 0.03 & 0.8 & 0.3 & V1\\
J0116$-$6153 & J0116.6$-$6153 & f & 01 16 19.70 & $-$61 53 43.00 & 0.500 & 0.799 & 33.50 &  & 30.50 & 0.30 &  & 0.60 & -0.04 & -0.30 & 0.05 & 0.20 & -0.19 & 0.04 & 2.7 & 1.2 & A1\\
J0143$-$5845 & J0143.6$-$5844 & f & 01 43 47.45 & $-$58 45 51.80 & 0.500 & 0.799 & 24.00 &  & 22.30 & 0.20 &  & 0.40 & -0.18 & -0.47 & 0.06 & 0.17 & -0.15 & 0.04 & 1.5 & 1.2 & A1\\
\hline
\end{tabular}
\footnotesize
  \begin{flushleft}
  Column description: (1) IAU conforming name; (2) Name listed in 2FGL catalog; 
                      (3) flags: e for extended source, f for flat spectrum source;
                      (4) right ascension coordinate of radio source;
                      (5) declination coordinate of radio source;
                      (6) error in right ascension; (7) error in declination;
                      (8) 5.0/5.5\,GHz flux density; (9) 7.3\,GHz flux density;
                      (10) 9.0\,GHz flux density; (11)-(13) corresponding flux 
                      density errors; (14) spectral index from 5.5\,GHz band;
                      (15) spectral index from 9.0\,GHz band only; (16)-(17), (19)
                      spectral index errors; (18) spectral index between 5.0/5.5 and
                      7.2/9.0\,GHz bands; (20) distance between radio and $\gamma$-ray 
                      localization; (21) normalized separation between radio and $\gamma$-ray localization, i.e.
ratio of the separation to the 1-$\sigma$ uncertainty; (22) observing campaign: V1 - VLA, A1/A2 - ATCA.
                      
  \end{flushleft}

  \end{table*}

  The NVSS catalog is derived from VLA observations at 1.4 GHz, the
SUMSS and MGPS-2 catalogs are derived from observations at 0.843 GHz
with the Molonglo Observatory Synthesis Telescope. These catalogs have
similar resolutions of $\sim 40''$ and are complementary to each
other, as they cover different areas of the sky. We searched for optical 
counterparts using the Supercosmos database \citep{2001MNRAS.326.1279H}. 
This database contains the digitized sky survey plates taken with the UK 
Schmidt telescope and is complete down to a B magnitude of B=22. Radio 
sources within $10^\circ$ of the Galactic plane (719 objects) were excluded in 
the optical cross-matching due to increased dust extinction and contamination 
of foreground stars at low Galactic latitudes. The ALLWISE catalog
includes point-like and resolved objects detected at infrared
wavelengths between 3.4 and 22\,$\mu$m with a corresponding angular
resolution between 6.1 and $12''$. We used a search radius of $20''$
to find counterparts in the NVSS, SUMSS, and MGPS-2 catalogs, $3.5''$ 
to search for optical counterparts and $7''$ for matching to an ALLWISE 
object. We obtained 687 matches with the ALLWISE catalog, 222 matches with the 
Supercosmos database, 610 matches with NVSS, 116 matches with SUMSS, and 161 
matches with MGPS-2. There were 47 objects that had both a counterpart in NVSS 
and MGPS-2 or SUMSS.

The radio spectral index was determined for all detected sources where
possible. For sources detected at both sub-bands independently, we
determined spectral indices between 5.0/5.5 and 7.3/9.0\,GHz directly
from their flux densities. For ATCA detections, the 2\,GHz bandwidths
at 5.5 and 9.0\,GHz were divided into 2048 spectral channels, allowing
determination of spectral indices even for objects that were found only in
one sub-band. No spectral index was determined for objects detected
only in one sub-band of a VLA observation since no multi-frequency
techniques were applied in the analysis due to the reduced bandwidth
per sub-band of 1\,GHz. As a note of caution, spectral indices
determined between both sub-bands in comparison to those
determined from only the low-sub-band do not necessarily match. The reason
for this can be systematic errors affecting the estimates in different
ways and intrinsic source spectra deviating from that of a power
law. For the following discussion we only include spectral indices of
sources that were detected in both sub-bands with either the VLA or
ATCA.

  \begin{table*}[thbp!]
  \centering
  \caption{The first 8 rows of 501 objects that were detected at 5.0/5.5 GHz and/or 
           7.3/9.0 GHz between 2.7 and 6.5 arcmin of the 2FGL counterpart localization.
           Table~\ref{tab:catI} is published in its entirety in the electronic
           edition. A portion is shown here for guidance regarding its form and
           content.\label{tab:catII}}
  \tiny\begin{tabular}{@{}c@{\,\,}c@{\,\,\,}c@{\,\,\,}c@{\,\,\,}c@{\,\,\,}r@{\,}r@{\,\,\,}r@{\,}r@{\,}r@{\,\,\,}c@{\,}c@{\,}c@{\,\,\,}c@{\,}c@{\,\,\,}c@{\,}c@{\,\,\,}c@{\,}c@{\,\,\,}c@{\,}c@{\,\,\,}c}
\hline
IAU name & 2FGL name & F & $\alpha$ & $\delta$ & $\Delta\alpha$ & $\Delta\delta$ & $F_5$ & $F_7 $ & $F_9$ & $\Delta F_5$ & $\Delta F_7$ & $\Delta F_9$ & $Sp_5$ & $Sp_9$ & $\Delta Sp_5$ & $\Delta Sp_9$ & $Sp$ & $\Delta Sp$ & $D$ & $N\sigma$ & Camp. \\
         &      &      & (h) (min) (s) & ($\circ$) (') ('') & ('') & ('') & (mJy) & (mJy) & (mJy) & (mJy) & (mJy) & (mJy) & & & & & & & (') & & \\ 
(1) & (2) & (3) & (4) & (5) & (6) & (7) & (8) & (9) & (10) & (11) & (12) & (13) & (14) & (15) & (16) & (17) & (18) & (19) & (20) & (21) & (22) \\ 
\hline
J0006$+$6823 & J0007.7$+$6825c &  & 00 06 35.75 & $+$68 23 20.16 & 0.050 & 0.050 & 14.07 &  &  &  &  &  &  &  &  &  &  &  & 6.4 & 1.7 & V1\\
J0014$-$0512 & J0014.3$-$0509  & f & 00 14 33.85 & $-$05 12 48.80 & 11.999 & 48.802 & 3.20 &  &  & 3.50 &  &  &  & 0.40 &  &  &  &  & 5.1 & 1.4 & A1\\
J0031$+$0724 & J0031.0$+$0724  & f & 00 31 19.71 & $+$07 24 53.61 & 0.050 & 0.050 & 6.42 & 5.88 &  & 0.25 & 0.32 &  &  &  &  &  & -0.23 & 0.18 & 3.4 & 1.5 & V1\\
J0039$+$4332 & J0039.1$+$4331  &  & 00 39 22.81 & $+$43 32 52.21 & 0.050 & 0.050 &  & 2.10 &  &  & 0.34 &  &  &  &  &  &  &  & 3.1 & 1.2 & V1\\
J0049$-$6344 & J0048.8$-$6347  & f & 00 49 32.92 & $-$63 44 11.39 & 2.531 & 4.723 & 1.70 &  &  & 0.40 &  &  & -3.00 &  & 1.29 &  &  &  & 5.7 & 2.0 & A2\\
J0116$-$6150 & J0116.6$-$6153  &  & 01 16 56.83 & $-$61 50 12.50 & 0.900 & 1.102 & 7.10 &  & 5.40 & 0.40 &  & 1.20 & -1.04 &  & 0.29 &  & -0.56 & 0.47 & 3.5 & 1.8 & A1\\
J0116$-$6156 & J0116.6$-$6153  & f & 01 16 43.97 & $-$61 56 53.30 & 0.601 & 0.799 & 17.50 &  & 15.80 & 0.40 &  & 1.30 & -0.31 &  & 0.14 &  & -0.21 & 0.17 & 3.7 & 1.9 & A1\\
J0152$+$8557 & J0158.6$+$8558  &  & 01 52 48.19 & $+$85 57 03.50 & 0.050 & 0.050 & 10.67 &  &  &  &  &  &  &  &  &  &  &  & 6.3 & 1.6 & V1\\
\hline
\end{tabular}
\footnotesize
  \begin{flushleft}
  Column description: (1) IAU conforming name; (2) Name listed in 2FGL catalog; 
                      (3) flags: e for extended source, f for flat spectrum source;
                      (4) right ascension coordinate of radio source;
                      (5) declination coordinate of radio source;
                      (6) error in right ascension; (7) error in declination;
                      (8) 5.0/5.5\,GHz flux density; (9) 7.3\,GHz flux density;
                      (10) 9.0\,GHz flux density; (11)-(13) corresponding flux 
                      density errors; (14) spectral index from 5.5\,GHz band;
                      (15) spectral index from 9.0\,GHz band only; (16)-(17), (19)
                      spectral index errors; (18) spectral index between 5.0/5.5 and
                      7.2/9.0\,GHz bands; (20) distance between radio and $\gamma$-ray 
                      localization; (21) normalized separation between radio and $\gamma$-ray localization, i.e.
ratio of the separation to the 1-$\sigma$ uncertainty; (22) observing campaign: V1 - VLA, A1/A2 - ATCA.
                      
  \end{flushleft}  

  \end{table*}

  \begin{table*}[htbp!]
  \centering
  \caption{The first 8 rows of 216 objects that were detected outside of the 6.5 arcmin and the were within
           the 99\% position uncertainty of the 2FGL counterpart localization.
           Table~\ref{tab:catIII} is published in its entirety in the electronic
           edition. A portion is shown here for guidance regarding its form and
           content.\label{tab:catIII}}
  \footnotesize\begin{tabular}{@{}c@{\,\,}c@{\,\,\,}c@{\,\,\,}c@{\,\,\,}c@{\,\,\,}c@{\,}c@{\,\,\,}r@{\,}r@{\,}r@{\,\,\,}r@{\,}c@{\,\,\,}c}
\hline
IAU name & 2FGL name & F & $\alpha$ & $\delta$ & $\Delta\alpha$ & $\Delta\delta$ & $F_5$ & $F_7 $ & $F_9$ & $D$ & $N\sigma$ & Camp. \\
         &      &      & (h) (min) (s) & ($\circ$) (') ('') & ('') & ('') & (mJy) & (mJy) & (mJy) & (') & & \\ 
(1) & (2) & (3) & (4) & (5) & (6) & (7) & (8) & (9) & (10) & (11) & (12) & (13) \\ 
\hline
J0001$+$6331 & J2358.9$+$6325 &  & 00 01 00.63 & $+$63 31 49.49 & 0.050 & 0.050 & 11.58 &  &  & 15.1 & 2.1 & V1\\
J0039$+$4336 & J0039.1$+$4331 &  & 00 39 28.71 & $+$43 36 51.13 & 0.050 & 0.050 & 19.92 &  &  & 6.5 & 2.5 & V1\\
J0047$-$634A & J0048.8$-$6347 & f & 00 47 40.59 & $-$63 47 39.37 & 0.990 & 1.768 & 3.40 &  & 2.20 & 7.9 & 2.7 & A2\\
J0103$+$1319 & J0103.8$+$1324 &  & 01 03 20.99 & $+$13 19 35.32 & 0.050 & 0.050 & 16.53 &  &  & 8.1 & 2.8 & V1\\
J0125$-$2325 & J0124.6$-$2322 & f & 01 25 02.21 & $-$23 25 52.90 & 0.270 & 0.241 & 13.50 &  & 7.40 & 6.9 & 2.8 & A2\\
J0126$+$6335 & J0128.0$+$6330 &  & 01 26 40.47 & $+$63 35 28.21 & 0.050 & 0.050 &  & 17.64 &  & 10.3 & 1.8 & V1\\
J0133$-$4414 & J0133.4$-$4408 & f & 01 33 06.40 & $-$44 14 21.41 & 0.068 & 0.083 & 45.70 &  & 41.60 & 6.9 & 2.9 & A2\\
J0149$+$8601 & J0158.6$+$8558 &  & 01 49 34.23 & $+$86 01 14.81 & 0.050 & 0.050 &  & 12.77 &  & 10.1 & 2.6 & V1\\
\hline
\end{tabular}
\footnotesize
  \begin{flushleft}
  Column description: (1) IAU conforming name; (2) Name listed in 2FGL catalog; 
                      (3) flags: e for extended source, f for flat spectrum source;
                      (4) right ascension coordinate of radio source;
                      (5) declination coordinate of radio source;
                      (6) error in right ascension; (7) error in declination;
                      (8) 5.0/5.5\,GHz flux density upper limit; (9) 7.3\,GHz flux density upper limit;
                      (10) 9.0\,GHz flux density upper limit; (11) distance between radio and $\gamma$-ray 
                      localization; (12) normalized separation between radio and $\gamma$-ray localization, i.e.
ratio of the separation to the 1-$\sigma$ uncertainty; (13) observing campaign: V1 - VLA, A1/A2 - ATCA.
                      
  \end{flushleft}    
  \end{table*}

\begin{table*}[bhtp!]
  \centering
  \caption{Possible counterparts from previous radio source 
           catalogs, the ALLWISE catalog, and the Supercosmos database for 
           radio sources found in Cat I, II, and III. Table~\ref{tab:counterparts} 
           is published in its entirety in the electronic
           edition. A portion is shown here for guidance regarding its form and
           content.\label{tab:counterparts}}
  \footnotesize\begin{tabular}{@{}c@{\,\,}r@{\,\,}c@{\,\,\,}c@{\,\,\,}c@{\,\,\,}c@{\,\,\,}c@{\,\,\,}c@{\,\,\,}c@{\,\,\,}c@{\,\,\,}c}
\hline
IAU name & $F_1$ & $C_1$ & $I_1$ & $I_w$ & $D_\mathrm{opt}$ & $\alpha_\mathrm{opt}$ & $\delta_\mathrm{opt}$ & B    & R     & $C_\mathrm{opt}$ \\
         & (mJy)   &       &       &       & ('')               &  (h) (min) (s)              & ($\circ$) (') ('')          & (Mag.) & (Mag.)  &                  \\ 
  (1)    & (2)   & (3)   & (4)   & (5)   & (6)              & (7)                   & (8)                   & (9)  & (10)  & (11) \\ 
\hline
J0001$+$6331 & 7.3 & J000100$+$633152 & N & J000100.72$+$633150.9 &  & - & - & &  &- \\
J0002$+$6219 & 12.7 & J000253$+$621917 & N & J000253.17$+$621917.6 &  & - & - & &  &- \\
J0004$+$2206 &  & - & - & J000407.36$+$220615.6 & 0.4 &00 04 07.39 & +22 06 15.7 &21.1 & 20.0 &2 \\
J0008$+$6837 & 264.4 & J000833$+$683721 & N & J000833.39$+$683722.1 &  & - & - & &  &- \\
J0014$-$0512 & 4.3 & J001433$-$051244 & N & J001433.87$-$051249.7 &  & - & - & &  &- \\
J0031$+$0724 & 11.6 & J003119$+$072456 & N & J003119.71$+$072453.4 & 0.5 &00 31 19.69 & +07 24 53.2 &19.3 & 19.1 &2 \\
J0031$-$5511 & 18.1 & J003156$-$551149 & S & J003156.87$-$551145.1 &  & - & - & &  &- \\
J0031$-$5521 & 381.2 & J003102$-$552107 & S & J003102.47$-$552104.3 &  & - & - & &  &- \\
\hline
\end{tabular}
\footnotesize
  \begin{flushleft}
  Column description: (1) IAU conforming name; (2) Radio counterpart flux density 
                      listed in NVSS/SUMSS/MGPS-2; (3) Name of counterpart NVSS/SUMSS/MGPS-2,
                      if there are matching counterparts found in multiple catalogs, the name
                      of the NVSS counterpart is listed; (4) catalogs in which radio counterparts
                      were found N $=$ NVSS, M $=$ MGPS-2, and S $=$ SUMSS; 
                      (5) Name of ALLWISE counterpart; (6) distance of an optical counterpart
                      from the $\gamma$-ray localization; (7),(8) position of optical counterpart right ascension
                      and declination; (9) optical B magnitude; (10) optical R magnitude;
                      (11) optical classification `1' - extended morphology, `2' - stellar or point-like object.
                      
  \end{flushleft} 
\end{table*}

The final list of 865 sources can be split into four different
categories:

(i) \textbf{Category I:} We detected 148 objects at arcsec scales in the 5.0/5.5 and
7.3/9.0\,GHz sub-bands within $2.7'$ of the localization of the
2FGL object, which also corresponds to the pointing direction of the
radio observations with the exception of fields observed using
mosaicking. Among those sources 125 were detected at both
sub-bands. For all sources in this category we provide the following
information in Table~\ref{tab:catI}: $\gamma$-ray source name; IAU
name of the detected radio source; J2000 coordinates followed by
1$\sigma$ uncertainties in arcsec; flux densities at 5.0/5.5 and
7.3/9.0\,GHz in mJy corrected for beam attenuation, followed by their
standard deviations, spectral indices, and distance of the source from the
pointing direction. Source flags are listed in column 3,
where `e' stand for a source that is extended and `f' is listed if the
source has a spectral index flatter than $-0.5$.

(ii) \textbf{Category II:} We detected 501 objects at arcsec scales in the 5.0/5.5\,GHz
and/or 7.3/9.0\,GHz sub-bands between 2.7 and $6.5'$ of the pointing
direction and within 3$\sigma$ of the $\gamma$-ray localization.
Similarly to Table~\ref{tab:catI} we list the Category II sources in
Table~\ref{tab:catII}. 

(iii) \textbf{Category III:} We detected 216 objects at arcsec scales
that were found beyond $6.5'$ from the 2FGL localization but were 
within the 3$\sigma$ confidence of the 2FGL $\gamma$-ray source 
localization. These objects are listed in Table~\ref{tab:catIII}. Flux 
density values provided in this table are to be considered lower limits 
due to beam attenuation, consequently no errors for those values are 
provided.

In addition we have detected 403 arcsec-scale sources outside of the 3$\sigma$
error ellipse of the 2FGL $\gamma$-ray source localization
(\textbf{Category IV}).  This list of sources was not tabulated since
they are not considered to be candidates for association with a
$\gamma$-ray source. A fill value of $-9.9$ in all tables indicates a
lack of information. In the case where there is a flux density listed
but the corresponding error is listed with a value of $-9.9$ then the flux 
density value is to be treated as a lower limit. The presence of additional associations for 2FGL
objects reported in the literature since the release of the 2FGL
catalog are marked with a flag. The convention for the
IAU names was applied to the list of all sources regardless of the
category defined above. It thus allows unique identification of radio
point sources using this identifier across all category tables. A
letter replaces the last digit of the IAU name if there were multiple
separate radio sources sharing the same name. All sources were
sorted in right ascension order, then names were checked for double
entries. If identical names were found, we replace the 10th digit with
a letter A. If the 10th letter of the previous source was A, then
change the name to B, and so forth.

In Table~\ref{tab:counterparts} we list multiwavelength counterpart information
for sources listed in Tables~\ref{tab:catI}, \ref{tab:catII}, and \ref{tab:catIII}
using IAU name as the unique identifier. If the source was associated with an object 
from either NVSS, SUMSS or MGPS-2 catalogs, its flux density at 1.4 GHz
(NVSS) or 0.843 GHz (SUMSS and MGPS-2) is listed along with the corresponding
source identifier in that catalog. In cases where both a NVSS and
SUMSS or MGPS-2 source is associated, the flux density value and name
of only the NVSS source is listed. In this case the presence of an
additional association over that of NVSS is indicated in a separate
column. If the source had an optical counterpart in Supercosmos, 
the optical position, B and R magnitude and classification is listed. 
This classification is based on the optical morphology where `1' represents an 
extended morphology and `2' indicates a stellar or point-like object.
If the source was associated with a WISE object, its WISE
source ID is listed as well. 

\begin{figure}[htbp!]
  \begin{center}
  \includegraphics[height=\columnwidth,angle=-90]{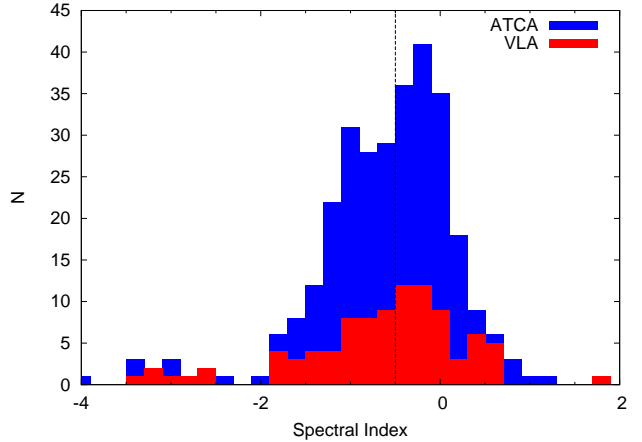}
  \end{center}
  \caption{Stacked histogram of radio spectral indices of category I and II objects found by ATCA and VLA
           between the frequencies 5.5 and 9.0\,GHz and 5.0 and 7.3\,GHz respectively. The vertical
           dashed line indicates the $-$0.5 value above which sources are considered flat spectrum.\label{fig:spix_dist}}
\end{figure}

\begin{figure}[htbp!]
  \begin{center}
  \includegraphics[height=\columnwidth,angle=-90]{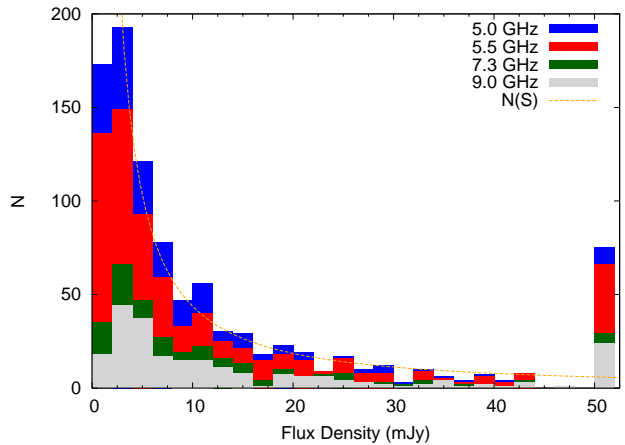}
  \end{center}
  \caption{Stacked histogram of the flux density distribution of all radio sources found within 6.5 arcmin 
           of the $\gamma$-ray localization (category I and II). The frequencies at 5.0 and 7.3 GHz represent the VLA observations, while
           5.5 and 9.0 GHz represent the data from ATCA observations. The last bin aggregates the number 
           of sources found above 50 mJy. The orange dashed line represents the expected flux density distribution
           of compact sources extrapolated and scaled from 8\,GHz observations of correlated flux density 
           of regions smaller than 50\,mas.\label{fig:flux_dist}}
\end{figure}

\begin{figure}[htbp!]
 \begin{center}
 \includegraphics[height=\columnwidth,angle=-90]{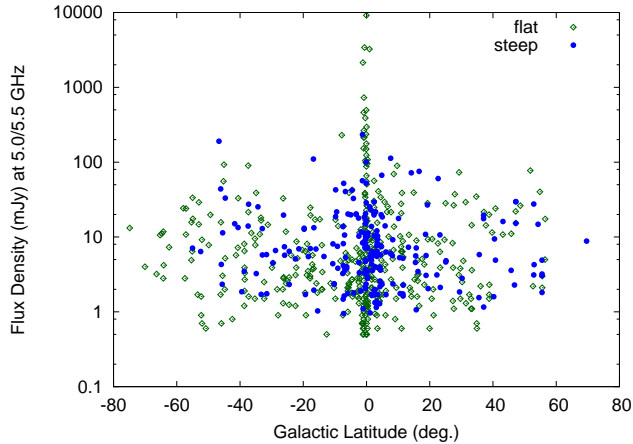}
 \end{center}
 \caption{Flux density dependence of detected sources at 5.0/5.5\,GHz in categories I and II with
          respect to Galactic latitude. Flat-spectrum ($\alpha > -0.5$) sources 
          are shown by green hollow squares and steep-spectrum sources 
          ($\alpha < -0.5$) are shown by filled blue circles.\label{fig:galactic_dist}}

\end{figure}

The spectral index of category I and II objects was determined for 304
objects and their distribution is visualized in Fig.~\ref{fig:spix_dist}. The
distribution is similar for both ATCA and VLA with a sharp rise in source counts
with spectral indices $>-2.0$. For Cat I objects 77 out of 137 are considered 
flat spectrum with a spectral index above $-$0.5. For the combined distribution
the split between flat and steep spectrum sources is almost exactly 50/50, with
153 steep spectrum and 151 flat spectrum sources. The flux density distribution 
of category I and II objects found is shown in 
Fig.~\ref{fig:flux_dist}. The source
counts drop slightly below 2.5\,mJy, thus given the expected source
count distribution we can assume the observations to be complete to a
flux density level of 2.5\,mJy. The ATCA observations find a factor of
four more sources at 5.5 and 9.0 GHz with flux densities above 50
mJy. This can be explained by the shallowness of existing
multifrequency catalogs in the southern hemisphere with respect to the
northern hemisphere. This causes many radio sources in the northern
hemisphere to be already associated with $\gamma$-ray sources
therefore creating a bias toward the southern hemisphere. The overall
flux density distribution is comparable to that of the extrapolated
and scaled log\,$N$--log\,$S$ distribution observed for compact sources
($\leq50$\,mas) found across the entire sky between 0.18 and 5.0\,Jy
at 8 GHz \citep[see also][]{2013MNRAS.432.1294P}.  No correlation was
found between radio flux density and spectral index for category I
objects.  The source distribution in Galactic latitude between flat
spectral index and other types of detected radio sources with a steep
or no spectral index at 5.0/5.5 GHz is comparable, see
Fig.~\ref{fig:galactic_dist}. The distribution of sources with respect
to Galactic latitude clearly shows an overabundance of detected
sources around the Galactic plane. Most
AGN surveys focus on regions outside of the Galactic plane thus
the Second Catalog of Active Galactic Nuclei detected by the Fermi Large Area 
Telescope requires Galactic latitudes $>\left|10\right|^\circ$ 
\citep{2011ApJ...743..171A}. Additionally, catalogs of other types of $\gamma$-ray
emitting sources, namely supernova remnants and pulsars have a bias toward the 
Galactic plane but neglect other parts of the sky. Looking at the distribution
of unassociated $\gamma$-ray sources reported in 2FGL and that were observed with
VLA and ATCA, we find that 54\% fall within $|b|<10^\circ$.

\subsection{Fields with no detections}

We found that 129 out of the 589 unassociated $\gamma$-ray sources
observed did not show a single compact radio source within their
3$\sigma$ confidence localization above the detection limits of our
observations of $\sim$1--2\,mJy. In 10 of those pulsed emission was
detected, one was related to emission from a Supernova Remnant, and
one was related to emission from a Low Mass X-ray binary (see
Section~\ref{sec:other_assoc}).  The positions and rms sensitivity
limits of the remaining 117 ``empty'' fields are listed in
Table~\ref{tab:nondetections}. No flag indicates that no point source
was detected in both sub-bands within the entire field of view, a `*'
flag indicates that at least one point source was detected outside of
the 3$\sigma$ 2FGL confidence localization error and `e' indicates the
possible presence of an extended source near the center of the
field. 

\begin{figure}[t]
    \centering
    \includegraphics[angle=-90,width=1.05\columnwidth,trim=0.8cm 0cm 0cm 0cm]{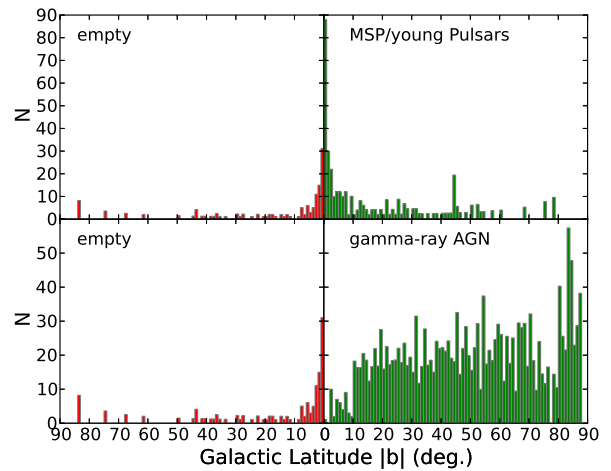}
    \caption{Distribution of sources in Galactic latitude $|b|$ of unassociated 
             2FGL sources for which no radio point source was found within the 
             3$\sigma$ confidence localization (red histograms on the left). 
             For comparison, the top right histogram in green shows the Galactic
             latitude distribution of millisecond pulsars ($p<0.01$\,s) and 
             young pulsars (age $<$ 20,000\,years) extracted from the ATNF pulsar 
             catalog \citep[v1.50;][]{2005AJ....129.1993M}.
             In the bottom-right panel the distribution of $\gamma$-ray loud AGN
             from 2LAC \citep{2011ApJ...743..171A} and new AGN associations made
             in this work is shown. The number counts were divided by 
             $\cos{(|b|)}$ to compensate for the change in sky area covered.\label{fig:empty_fields}}

\end{figure}

The distribution of the empty fields in Galactic latitude $|b|$ is
shown in Fig.~\ref{fig:empty_fields} together with the
distributions of $\gamma$-ray loud AGN and the known class of
$\gamma$-ray loud pulsars, namely millisecond pulsars (pulse period
$<0.01$\,s) and young pulsars (age $< 20,000$\,years).  Most of the
empty fields are found along the Galactic plane with 25\% falling
within $\pm1^\circ$ and 67\% falling within $\pm10^\circ$ of the
Galactic plane.  The 1-100\,GeV $\gamma$-ray fluxes of the empty
fields range between $2.3\cdot10^{-10}$ and
$1.6\cdot10^{-8}$\,ph\,cm$^{-2}$\,s$^{-1}$ with a mean value of
$2.7\cdot10^{-9}$\,ph\,cm$^{-2}$\,s$^{-1}$. The average $\gamma$-ray
photon spectral index is 2.34$\pm$0.29 and only two fields had a
photon index of less than 1.8. No correlation was found between the
$\gamma$-ray fluxes and photon indices of empty fields. The general
population of unassociated $\gamma$-ray sources from 2FGL has 67\%
concentrated at $|b|<10^\circ$ and 26\% at $|b|<1^\circ$. Comparing
the distribution of empty fields in Galactic latitude with that of the
known $\gamma$-ray AGN population we find no clear correspondence
between these two. However, when comparing the empty fields to the
distribution of millisecond and young pulsars, we find a p-value from
the two-sided Kolmogorov-Smirnov statistical test of 0.21, which
indicates that both distributions could be drawn from the same
population. We should note that the empty fields are
significantly more concentrated in the Galactic plane, which indicates
the presence of a Galactic population of unassociated $\gamma$-ray
sources with a distribution similar to that of millisecond and young pulsars. 
No optical or IR localization of a $\gamma$-ray source in the empty
fields is possible since their localization errors are in a range of arcminutes.

Given the lack of compact radio emission in these 117 2FGL empty 
fields, we searched existing single
dish surveys for large scale radio emission, mainly the Effelsberg
21cm and 11cm surveys focusing on the Galactic plane
\citep{1990AAS...85..805F,1997AAS..126..413R}\footnote{Image data is
  available at: \url{http://www3.mpifr-bonn.mpg.de/survey.html}}. Among
the 44 fields that are covered by the Effelsberg surveys, we found 33 fields
which had extended structures at or near the $\gamma$-ray localization. In
many cases structures resembled supernova remnant shells (bubbles). A small 
representative selection of the structures found in the 11\,cm Effelsberg 
survey is shown in Fig.~\ref{fig:extended_sources}.  A qualitative description 
of those is given below:
\begin{itemize}
 \item 2FGL\,J0225.9+6154c: A large scale structure resembling a figure of eight
with a bright core is found. The $\gamma$-ray source is located near the center
of this structure. Overall, the structure resembles that of a bipolar outflow around a hot core.

 \item 2FGL\,J1746.6-2851c: The $\gamma$-ray localization falls onto an arc-like structure
                            resembling that of a supernova remnant shell. 
       
 \item 2FGL\,J1924.8+1724c: This $\gamma$-ray source lies very close to the innermost 
                            region of the Galactic plane. The $\gamma$-ray source falls
                            onto an extended radio source of yet unknown origin.

 \item 2FGL\,J2019.1+4040: The $\gamma$-ray localization falls onto the Northern edge of a 
                           sphere-like structure resembling that of a supernova remnant shell. 
\end{itemize}

We also searched the catalog of Galactic supernova remnants (SNR) with
radio emission compiled by \citep{2014arXiv1409.0637G}
that lists 274 known objects for possible matches against the 
empty fields. We used the 
known size listed in the Green catalog, and in cases where there is no reliable
size used a cone search radius of 0.2$^\circ$. We found 13 matches
among the empty fields leaving still a large fraction of the empty
fields spatially unrelated to known SNR. A listing of
the matches can be found in Section~\ref{sec:other_assoc}. Further
investigation of the empty fields that could confirm the nature of the
radio structures found is anticipated in the future.

\begin{table*}[htbp!]
  \small\centering
  \begin{center}
 \caption{The first 8 rows of 117 2FGL fields for which no radio point source 
          was detected within the 4$\sigma$ localization error at both 5.0/5.5 
          and 7.3/9.0 GHz. Table~\ref{tab:nondetections} is published in its entirety in the electronic
           edition. A portion is shown here for guidance regarding its form and
           content. \label{tab:nondetections}}
 \begin{tabular}{lllllllrrcc}
\hline
2FGL & \multicolumn{3}{c}{Right Ascension} & \multicolumn{3}{c}{Declination} & l & b & rms & flag\\
     & \multicolumn{3}{c}{(h) (min) (s)}         & \multicolumn{3}{c}{($^\circ$) (') ('')} & ($^\circ$) & ($^\circ$) & (mJy beam$^{-1})$ &\\
\hline
J0032.7$-$5521  & 00 & 32 & 43 & $-$55 & 21 & 22 & 308.5234 & $-$61.5682 & 0.15/0.19 & *\\
J0038.8$+$6259  & 00 & 38 & 53 & $+$62 & 59 & 48 & 121.5072 & $+$0.1597   & 0.083/0.080 & *\\
J0129.4$+$2618  & 01 & 29 & 29 & $+$26 & 18 & 35 & 133.4507 & $-$35.7836 & 0.073/0.068 & *\\
J0212.1$+$5318  & 02 & 12 & 09 & $+$53 & 18 & 19 & 134.9371 & $-$7.6745  & 0.076/0.074 &\\
J0214.5$+$6251c & 02 & 14 & 33 & $+$62 & 51 & 11 & 132.2501 & $+$1.4947   & 0.078/0.076 & *\\
J0218.7$+$6208c & 02 & 18 & 43 & $+$62 & 08 & 22 & 132.9366 & $+$0.9752   & 0.087/0.083 & *\\
J0225.9$+$6154c & 02 & 25 & 59 & $+$61 & 54 & 12 & 133.8172 & $+$1.0463   & 0.098/0.089 & *\\
J0251.0$+$2557  & 02 & 51 & 03 & $+$25 & 57 & 57 & 153.9615 & $-$29.6029 & 0.085/0.078 & *\\
\hline 
\end{tabular}

 \begin{flushleft}\footnotesize
 Note: The rms is given for both the 5.5/5.5 and 7.3/9.0 GHz observations. The
       `*' flag indicates whether a point source outside of the 4$\sigma$ 
       $\gamma$-ray localization was found. The `e' flag indicates whether
       evidence for an extended source was found in the field.
 \end{flushleft}
  \end{center}
 \end{table*}

\begin{figure*}[htbp!]
  \begin{center}
    \includegraphics[width=0.8\textwidth]{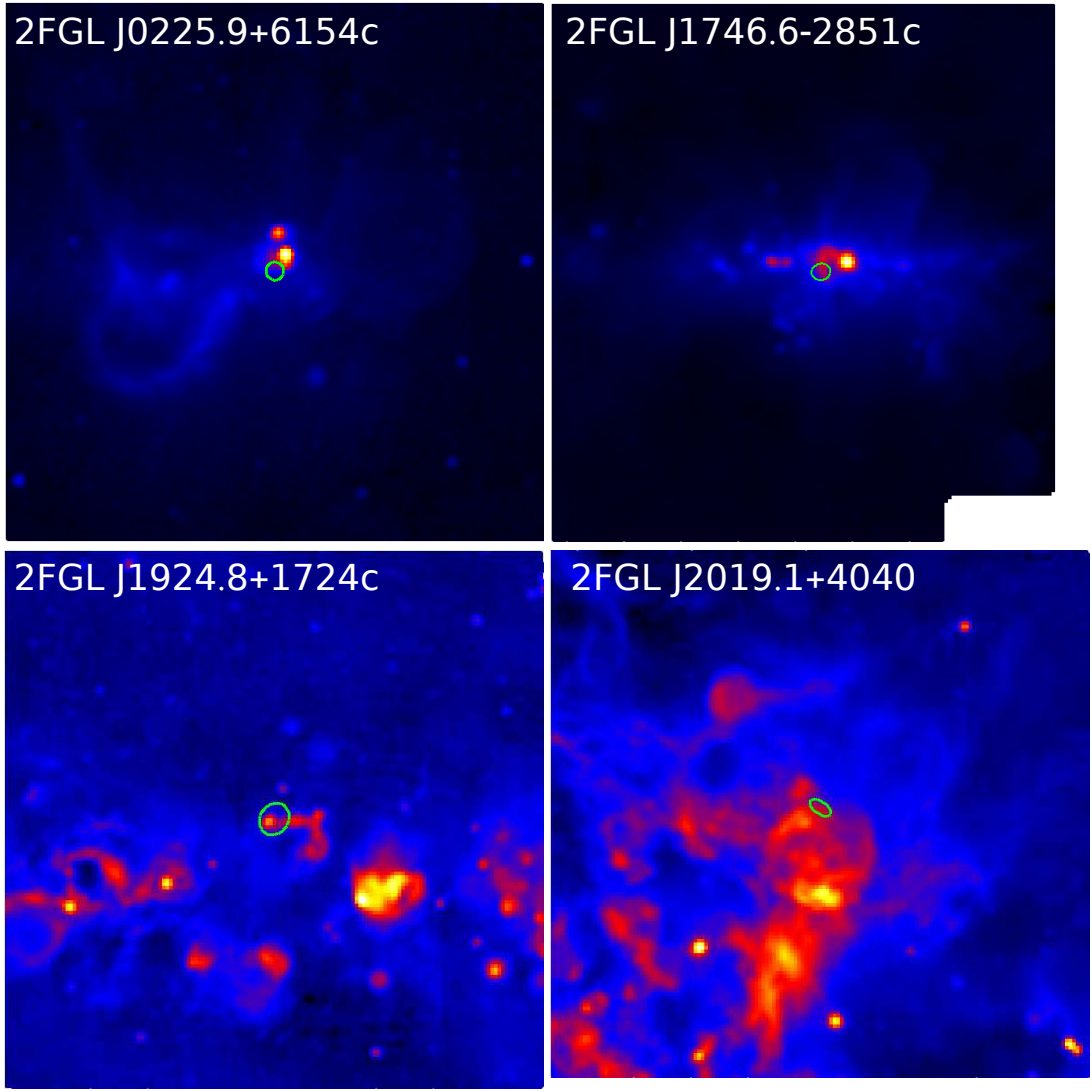}
  \end{center}
  \caption{Example of four ``empty'' fields that were observed in the 
           Effelsberg 100m telescope 11cm radio continuum survey of the 
           Galactic plane \citep{1997AAS..126..413R,1990AAS...85..805F}.
           The images cover a sky area of 5x5 degrees centered on the
           four 2FGL sources J0225.9+6154c, J1746.6$-$2851c, J1924.8+1724c, 
           and J2019.1+4040 respectively. The green circle indicates the 3$\sigma$
           localization of the $\gamma$-ray point source. The maps are shown
           in Galactic coordinates (l/b) with no projection applied. In the 
           case of 2FGL J1746.6$-$2851c not the entire 5x5 degree field was covered
           by the survey.\label{fig:extended_sources}}
\end{figure*}

\subsection{Associations with Compact Sources}\label{sec:3.3}

Follow-up VLBI observations of 170 selected
candidate targets obtained from the VLA and ATCA observations resulted
in the detection of compact parsec-scale emission in 95 of the
targets. The success rate of detection of parsec-scale emission for
sources detected by ATCA was 54\% and for VLA selected targets was
65\%. To reliably associate a $\gamma$-ray source with a radio source
that exhibits parsec-scale radio emission, we use the fact that compact radio 
sources are relatively rare objects. Their number increases with a decreasing
flux density limit. Fitting an empirical relationship 
for the source counts ($\log N$--$\log S$
diagram), we obtain the expected number of compact radio sources
with flux densities greater than a given value. Under the assumption
of an isotropic distribution of radio sources, we can evaluate the
probability of finding a background compact source brighter than a
given flux density within the distance $d$ between a $\gamma$-ray
source and its possible radio counterpart. Using this probability, we
compute the likelihood ratio of association defined as the
probability that the radio and $\gamma$-ray sources found at a
distance $d$ is physically the same object (and their position
difference is due to statistical errors only), $P_1$, to the
probability that the radio source is a background, unrelated object,
$P_2$. The first probability is determined as
\begin{equation}
  P_1 = e^{-n^2/2}
\end{equation}
  where $n$ is the normalized distance between the radio and $\gamma$-ray source.
For the case of a circular error ellipse, $n = d/\sigma$. In the general case

\begin{equation}
  n = \sqrt{\Delta^2\alpha \cos^2 \delta + \Delta^2\delta} \:
      \Frac{\sqrt{\sigma^2_{\rm maj} \sin^2 \beta + \sigma^2_{\rm min} \cos^2 \beta}}
           {\sigma_{\rm maj} \, \sigma_{\rm min} },
\end{equation}
   where angle $\beta$ is 
\begin{equation}
  \beta = \arctan{\Frac{\Delta\delta}{\Delta\alpha \, \cos\delta} - (\pi/2 - \theta)}
\end{equation}
  and $\sigma_{\rm maj}, \sigma_{\rm min}$ are semi-major and semi-minor axes
of the error ellipse, and $\theta$ is rotation angle of the error ellipse.

Our analysis of the normalized distribution of 1081 arc lengths between
reported \textit{Fermi} positions and VLBI positions revealed that the reported
parameters of the error ellipse axes should be scaled by a factor of 0.897. We 
used a rescaled $\sigma_{\rm maj}, \sigma_{\rm min}$ in our analysis. 
Remarkably, in 2FGL \citep{2012ApJS..199...31N}, the reported uncertainties were
artificially multiplied by a factor of 1.1. Our analysis shows that applying 
the 1.1 scale factor was an oversight and should be undone.

  If we consider the position of a given $\gamma$-ray source is precisely known, then the probability
to find a background radio source with a given flux density at distance $d$ or closer 
is equal to the product of the total number of sources inside an area with
radius $d$ to the total area of the celestial sphere, i.e.
\begin{equation}
  P_2 = N(F) \, \frac{d^2}{4}.
  \label{e:p2}
\end{equation}

According to \citet{2013MNRAS.432.1294P}, analysis of the complete sample
of VLBI detected sources allows us to approximate $N(F)$ as 
$ 327 \cdot F^{-1.237}$, where $F$ is the correlated flux density in Jy
at 8~GHz from regions of milliarcsecond size.

If we take into account that the position of a $\gamma$-ray source is not
precisely known and the probability for a $\gamma$-ray photon to be found at 
a certain location can be described by a two-dimensional Gaussian distribution,
the probability $P_2$ is expressed via the integral
\begin{equation}
  \begin{array}{lcl}
  P_2 & = & N(F) \Frac{1}{4\,\pi \, \sigma_{\rm maj} \sigma_{\rm min}} \\
  & & 
  \displaystyle \int\!\!\int\limits_{-\infty}^{+\infty}
        G(\alpha,\delta,\alpha_0,\delta_0,\sigma_{\rm maj},\sigma_{\rm min})\\
   & &   \qquad\qquad  (\Delta^2\alpha \cos^2 \delta + \Delta^2\delta) 
        \, \cos\delta \, d\alpha d\delta,

  \end{array}
  \label{e:p2a}
\end{equation}
  where $G(\alpha,\delta,\alpha_0,\delta_0,\sigma_{\rm maj},\sigma_{\rm min})$ is 
the Gaussian distribution:
\begin{equation}
  \begin{array}{l}
    G(\alpha,\delta,\alpha_0,\delta_0,\sigma_{\rm maj},\sigma_{\rm min}) = \\
    \qquad\exp\Biggl\{-\Frac{(\Delta\alpha\cos\delta\cos\theta - \Delta\delta\sin\theta)^2}{\sigma_{\rm maj}^2}\\
    \qquad\qquad\qquad - \Frac{(\Delta\alpha\cos\delta\sin\theta + \Delta\delta\cos\theta)^2}{\sigma_{\rm min}^2}\Biggr\}
  \end{array}
  \label{e:p2b}.
\end{equation}

Expression \ref{e:p2a} generalizes equation \ref{e:p2}. The probability
evaluated with expression \ref{e:p2a} is usually greater than evaluated 
with expression \ref{e:p2}.

  The likelihood ratio is just
\begin{equation}
   \Lambda = \Frac{P_1}{P_2}.
   \label{e:lik}
\end{equation}

\begin{table*}[tb!]
  \begin{center}
 \caption{The first 8 rows of 95 VLBI detected radio sources in the fields of
 2FGL unassociated $\gamma$-ray objects. The full table is available in the 
 supporting documents of the online version of this article. 
 Table~\ref{tab:vlbi} is published in its entirety in the electronic
 edition. A portion is shown here for guidance regarding its form and
 content.\label{tab:vlbi}}
 \begin{tabular}{lllllllllllr}
\hline
IAU Name & 2FGL & \multicolumn{3}{c}{Right Ascension} & \multicolumn{3}{c}{Declination} & S     & D       & N$\sigma$ & $\Lambda$\\
         &      & (h)     & (min) & (s)               & ($^\circ$) & ($'$) & ($''$)     & (Jy)  & ($''$)  &           &          \\
\hline
J0031$+$0724 & J0031.0$+$0724 & 00 & 31 & 19.7097 & $+$07 & 24 & 53.558 & 0.0130 & 3.40 & 1.29 & 25.2\\
J0039$+$4330 & J0039.1$+$4331 & 00 & 39 & 08.1595 & $+$43 & 30 & 14.619 & 0.0110 & 1.39 & 0.55 & 244.5\\
J0102$+$0944 & J0102.2$+$0943 & 01 & 02 & 17.1123 & $+$09 & 44 & 09.586 & 0.0210 & 1.24 & 0.47 & 706.2\\
J0103$+$1323 & J0103.8$+$1324 & 01 & 03 & 45.7410 & $+$13 & 23 & 45.258 & 0.0330 & 0.78 & 0.25 & 3366.0\\
J0116$-$6153 & J0116.6$-$6153 & 01 & 16 & 19.6126 & $-$61 & 53 & 43.514 & 0.0260 & 2.68 & 1.25 & 100.6\\
J0157$+$8557 & J0158.6$+$8558 & 01 & 57 & 03.8159 & $+$85 & 57 & 38.873 & 0.0050 & 1.75 & 0.45 & 60.9\\
J0221$+$2514 & J0221.2$+$2516 & 02 & 21 & 26.9651 & $+$25 & 14 & 33.665 & 0.0220 & 2.82 & 0.87 & 111.0\\
J0223$+$6821 & J0222.7$+$6820 & 02 & 23 & 04.5374 & $+$68 & 21 & 54.995 & 0.0150 & 1.97 & 1.13 & 109.2\\
\hline
\end{tabular}

  \end{center}
 \begin{flushleft}\footnotesize
     Column description:
     IAU Name --- IAU conforming name; 2FGL --- 2FGL identifying name; 
     Right Ascension/Declination --- J2000 coordinates of VLBI detection;
     S --- VLBI flux density; D --- separation between radio and $\gamma$-ray
     source; N$\sigma$ --- normalized separation between radio and $\gamma$-ray 
     source; $\Lambda$ --- likelihood ratio   
 \end{flushleft}
 \end{table*}

For 80\% of the 2FGL objects the $1\sigma$ position error is smaller
than $3.6'$. It follows from Eq.~\ref{e:lik} that for these sources if
we find a compact radio source brighter than 12~mJy at 8~GHz within
their 2FGL $2\sigma$ error ellipse (semi-major axis $7.2'$), the
likelihood ratio of their association is greater than 10. This
estimate shows that high resolution radio observations are a very
powerful method for establishing an association between radio and
$\gamma$-ray AGN. The VLBI positions of the detected compact sources,
the associated 2FGL field, and their association probability are
listed in Table~\ref{tab:vlbi}. The likelihood ratio shows how much
more a true association is likely than spurious association with a
background source.  For example a value of $\Lambda = 30$ indicates that
the probability
that the $\gamma$-ray and radio sources are physically related is a
factor of 30 greater than the probability they are unrelated. In this
case there is a 30:1 chance that it is correct to identify the
$\gamma$-ray source with the radio source. We accept a likelihood
ratio of better than 8.0 as sufficient to claim association of the radio
source with the respective $\gamma$-ray object. This value was chosen
from the distribution of obtained likelihood values. There are 19 detections
with a likelihood ratio of less than 8. None had a likelihood ratio between
8.0 and 10.0. Based on this criterion, we have found and
associated 76 new $\gamma$-ray loud AGN. In the four cases of
2FGL\,J0253.9+5908, 2FGL\,J0409.8-0357, 2FGL\,J1511.8-0513, and
2FGL\,J1844.3+1548 two high likelihood quasar counterparts were found
in each. It is conceivable that both sources
contribute to the observed $\gamma$-ray emission. We should note that
at least one element of these pairs is weak, 11--14~mJy, and the distance between objects
is 4--$5'$. According to Eqn.~\ref{e:p2}, the
probability to find a background source 12~mJy or brighter in a disk
of $5'$ radius is 0.041. Using the binomial distribution, we can
evaluate the probability of finding 4 pairs in a random sample of 76
objects: 0.20. This is a fairly high probability. Appearance of
multiple associations highlights the nature of the proposed
association method: it is statistical, not causal. When we consider
a specific object, the probability of false association is always
non-zero. But when we consider the population as a whole, we can
tolerate a controllable level of spurious associations.  In the
framework of our method we are in a position to evaluate
quantitatively the probability of such a contamination. For the
following analysis and discussion we have excluded these four objects
that have two counterparts each.

\begin{figure}[htbp!]
  \centering
  \includegraphics[width=0.7\columnwidth,angle=-90]{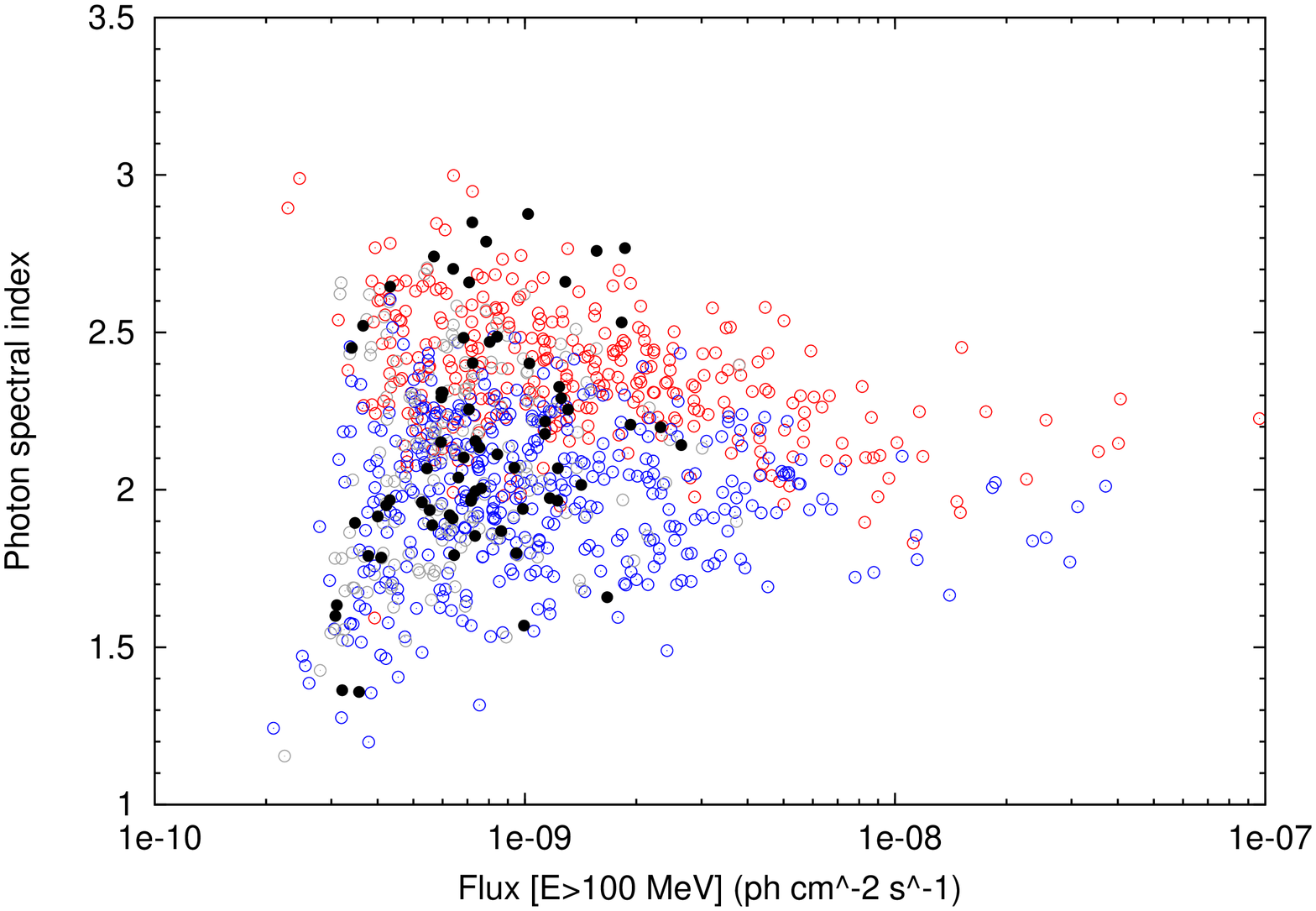}
  \includegraphics[width=0.7\columnwidth,angle=-90]{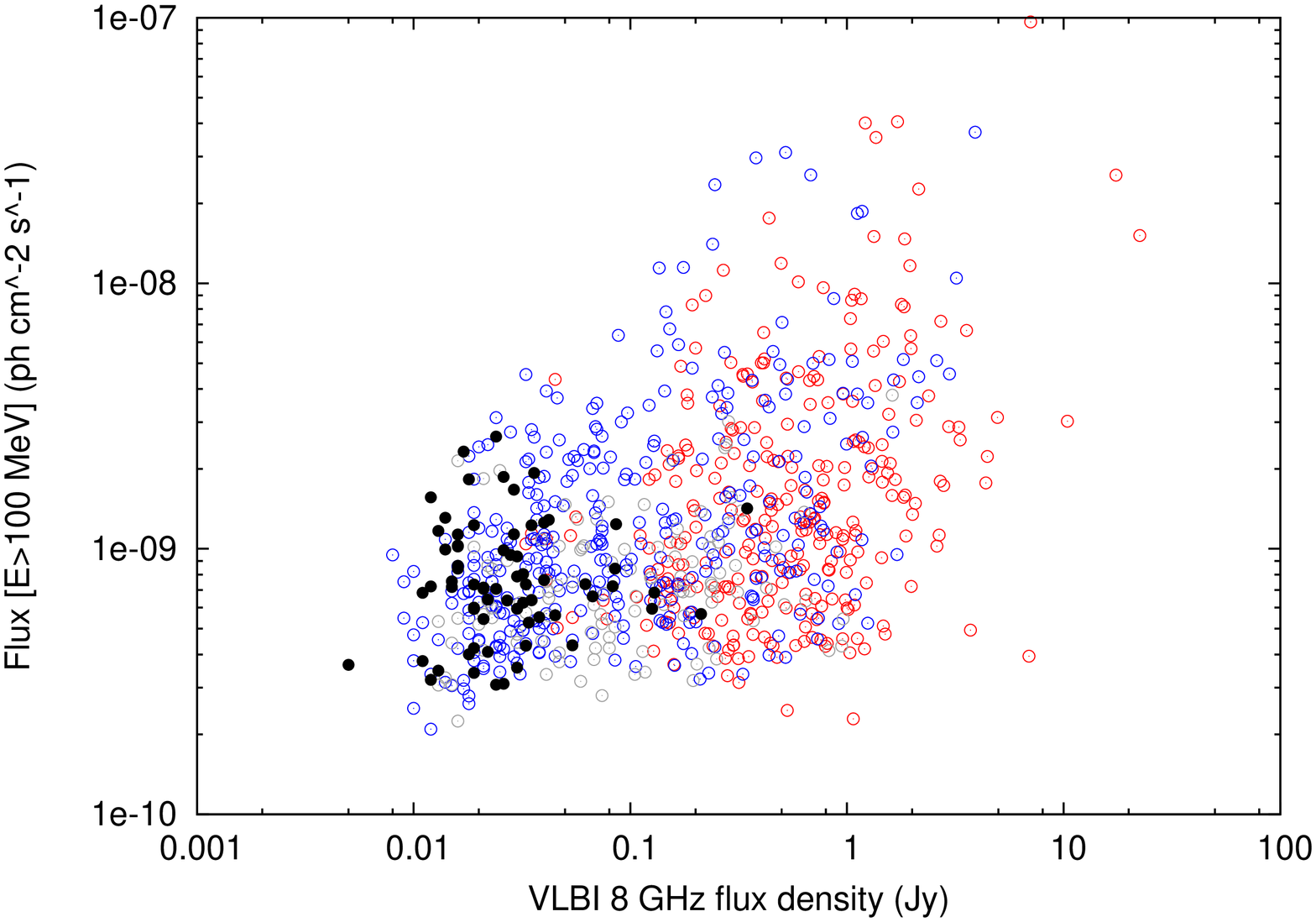}
  \caption{\textit{Top:} Photon spectral index plotted against the $\gamma$-ray
                  flux above 100 MeV for blazars in the clean 2LAC sample (empty circles)
                  and newly associated blazars (filled black circles). Source
                  classes from the clean sample are indicated by colors, red: FSRQs,
                  blue: BL Lac, grey: AGN of other type. \textit{Bottom:} Distribution
                  of VLBI 8 GHz flux densities of blazars from the clean 2LAC
                  sample (empty circles) and new identifications (black filled circles).
                  The colors are the same as on the left, based on blazar class. \label{fig:vlbi}}
\end{figure}

We now compare this new group of radio faint $\gamma$-ray
loud AGN to the previously known population of $\gamma$-ray loud AGN
listed as a clean sample in \citet[][2LAC]{2011ApJ...743..171A}. The
newly associated $\gamma$-ray loud AGN constitutes 8\% of the total
number of $\gamma$-ray associations obtained by combining the 2LAC
clean sample with our new detections. In Fig.~\ref{fig:vlbi} the
$\gamma$-ray flux above 100\,MeV is plotted against the $\gamma$-ray
photon index and against the 8 GHz VLBI flux density where
available\footnote{For the 2LAC clean sample the rfc-2014b catalog
  (\url{http://astrogeo.org/rfc}) was searched using the likelihood
  ratio for association of $\gamma$-ray counterparts with VLBI
  detections at the 8.0-8.8 GHz band only.} for blazars listed in the 2LAC clean
sample and the newly associated $\gamma$-ray loud blazars. The newly
associated $\gamma$-ray loud blazars match very well the general
distribution of blazars in the 2LAC clean sample adding to the fainter
end of $\gamma$-ray loud blazars that are predominantly BL Lac type
blazars. This confirms that the newly associated sources are
consistent with the previously found general population of
$\gamma$-ray loud blazars. There are only two 
new Fermi-AGN associations from our analysis with a flux density above 
150\,mJy, J2131.0-5417 and J0307.4+4915. This confirms an almost complete 
sample down to these levels of flux densities. This means that almost no 
\textit{Fermi} AGN identification with VLBI flux densities greater 
than 150--200\,mJy was previously missed.   

\subsection{Other Associations}\label{sec:other_assoc}

Since the release of the 2FGL catalog in 2011 other groups have also
attempted to find associations for sources listed as unassociated
$\gamma$-ray objects.  We performed a literature search for
unassociated 2FGL objects in order to find references on previously
reported associations or association attempts. We found that 24
unassociated objects are now associated with Pulsars, 2 are now
associated with AGN, and 38 are tentatively associated with or candidates for
association with supernova remnants, and two are associated with
binary star systems. The corresponding 2FGL identifier, association
names, and references are listed in Table~\ref{tab:assoc}. The
$\gamma$-ray source 2FGL\,J2339.6-0532 was associated with a
millisecond pulsar black widow type system, however to date no pulsed
$\gamma$-ray or radio emission has been detected from this pulsar
candidate.  Thus it is listed in Table~\ref{tab:assoc} with a `?' as
source type together with the relevant reference describing the nature
of the possible counterpart at other wavelengths. We also list
potential association candidates with Supernova Remnants by
cross-matching the list of unassociated sources against the catalog of
known Galactic supernova remnants \citep{2014arXiv1409.0637G}. These
entries were marked `SNR?' for their source type. The cross-matching
criteria are based on a spatial match, with the $\gamma$-ray
localization falling within the extent of a listed supernova remnant.
If no size is known or the size was marked with a `?' we assumed a
size of the SNR of $0.2^\circ$. We thus found 96 2FGL sources matching
this criteria with 51 unique entries in the Green catalog.

\begin{table}[t!]
  \caption{First 8 entries of 167 possible associations of 2FGL unassociated sources from literature.
         Table~\ref{tab:assoc} is published in its entirety in the electronic edition. A portion is 
         shown here for guidance regarding its form and content. The source types are: SNR = Supernova 
         Remnant; PSR = Pulsar; XSS = X-ray Binary Star System; BSS = Binary Star System.
  \label{tab:assoc}}

\begin{tabular}{llll}
\hline
2FGL & Association & Type & Reference \\
\hline
J0002.7+6220 & G116.9+0.2 & SNR? & \citep{2009BASI...37...45G}\\
J0106.5+4854 & PSR\,J0106+4855 & PSR & \citep{2012ApJ...744..105P}\\
J0128.0+6330 & G126.2+1.6 & SNR? & \citep{2009BASI...37...45G}\\
J0128.0+6330 & G127.1+0.5 & SNR? & \citep{2009BASI...37...45G}\\
J0214.5+6251c & G132.7+1.3 & SNR? & \citep{2009BASI...37...45G}\\
J0218.7+6208c & G132.7+1.3 & SNR? & \citep{2009BASI...37...45G}\\
J0221.4+6257c & G132.7+1.3 & SNR? & \citep{2009BASI...37...45G}\\
J0224.0+6204 & G132.7+1.3 & SNR? & \citep{2009BASI...37...45G}\\
\hline
\end{tabular}

 \end{table}

\section{Discussion}\label{sec:4}

\begin{figure}[t!]
  \centering
  \includegraphics[width=0.7\columnwidth,angle=-90]{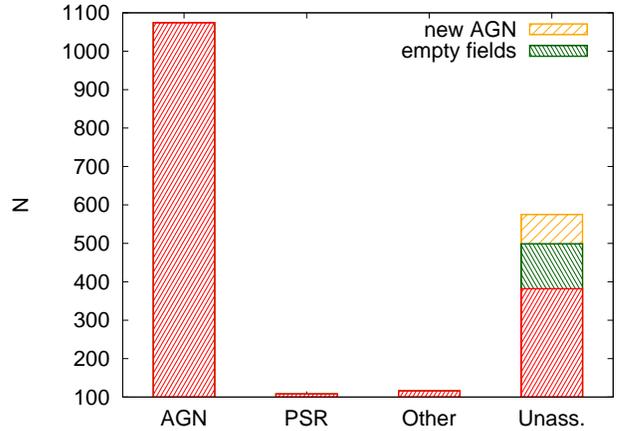}
  \caption{The total number of sources listed in 2FGL is 1873, 
           which can be broken down into 1074 AGN, 108 Pulsars (PSR), and 
           116 other types of sources, e.g. supernova remnants, etc. The
           2FGL also listed 575 unassociated sources, for which we found
           new AGN associations in 76 cases (13\% of the unassociated sources).
           The 117 ``empty'' fields without association listed make up 20\% of the 
           unassociated sources. Since the release of the catalog 28 new
           associations were reported among the unassociated $\gamma$-ray sources.
           \label{fig:chart_end}}
  
\end{figure}

Our multi-tiered approach observing the localization
error circles of unassociated $\gamma$-ray sources using radio interferometers
led to the association of 76 unassociated $\gamma$-ray sources with AGN, 
reducing the fraction of unassociated sources from 2FGL
to 27\%. Compared to other association attempts since the release
of 2FGL this makes our association method the most successful even compared to 
searches of pulsed emission among unassociated $\gamma$-ray sources. We expect
to increase the number of associations with fainter detections of sub-arcsec 
emission by applying phase referencing calibration on the already obtained data
and additional VLBI observations of candidates for association. Additional associations can be expected from 
planned VLBI observations of Category I sources. Due to selection effects we 
found significantly more new AGN within $|b|<10^\circ$ than what is typical for 
a uniform distribution of AGN around the sky. This is no surprise, since the 
focus of most of the association work regarding AGN has been on sky 
regions of $|b|>30^\circ$ and declinations of $>10^\circ$. In Figure~\ref{fig:chart_end} 
the major contributing source classes of identified or associated objects are shown in comparison to 
the number of unassociated sources, with the newly associated AGN shown in a 
separate color. Although strictly speaking an AGN is not the only type of object
with emission at 8 GHz from regions of milliarcsecond scale, the probability of 
contamination of our sample of associated sources with objects of another type is 
very low. There are no star-burst galaxies with a compact component brighter than 10\,mJy
detected. To date, only ten radio stars brighter than 10\,mJy are known out of the estimated 
total number of objects with emission from regions of milliarcsecond scales of 100,000. 

Radio searches for counterparts at low Galactic latitudes are complicated by 
the fact that the Galactic plane is bright at radio wavelengths, with contributions from
both diffuse emission and bright Galactic sources. For ATCA observations a 
significant fraction of search regions at low Galactic latitudes were found to 
have image rms noise levels considerably greater than those of fields at 
high Galactic latitude. The majority of these were found to be fields in close 
proximity to Galactic HII regions. At cm wavelengths, HII regions can be very 
bright (peak brightnesses of tens of Jy beam$^{-1}$) and as much as 5--$10'$ in 
extent, and so the $\gamma$-ray search field can be contaminated by radio 
emission from an HII region in the primary beam or one of the side-lobes. 
\citet{2003AA...397..213P} compiled a catalog of 1442 Galactic HII regions from 
previously published lists, and the number of known HII regions has been further 
expanded by \citet{2005AJ....129..348G}, with the addition of more compact regions,
and \citet{2014ApJS..212....1A}, among others. The presence of such sources 
near search fields makes searches for faint radio counterparts more difficult. 
A thorough analysis of HII regions in this context is anticipated in the future.

\begin{figure*}[htbp!]
   \centering
   \includegraphics[width=12cm]{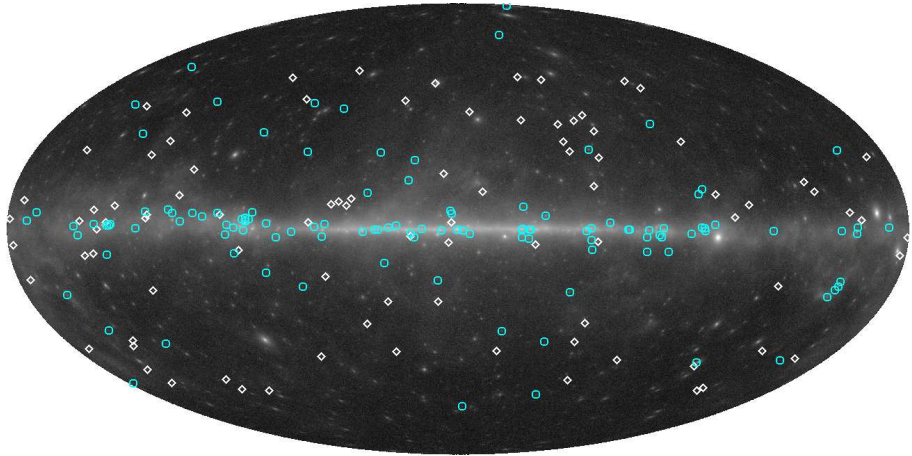}
   \caption{The \textit{Fermi} two-year all-sky map for energies above 1\,GeV
           (Credit: NASA/DOE/Fermi LAT Collaboration). The
           positions of empty fields are marked with cyan circles and the 
           positions of newly associated AGN are marked with white diamonds.\label{fig:allsky}}
 \end{figure*}

We find that a significant fraction of radio observations around
unassociated $\gamma$-ray sources resulted in non-detections of a
single radio source within 3$\sigma$ of the localization down to a
detection limit of $\sim$1\,mJy. These empty fields were shown to
be concentrated in the Galactic plane,
especially within the inner $2^\circ$ in Galactic latitude (see
Fig.~\ref{fig:allsky}). The spatial distribution of empty fields is
compatible with that of young pulsars and milli-second pulsars, which
are known to produce $\gamma$-ray emission. A look at existing single
dish survey maps revealed that 75\% of those empty
fields covered by the Effelsberg Galactic plane survey show extended radio 
emission with some of them resembling
supernova remnant shells. We searched the catalog of Galactic
SNRs \citep{2014arXiv1409.0637G} and found only 13 matches, while a
search over all unassociated $\gamma$-ray sources resulted in 96
spatial matches with known SNRs.  Thus, there
seems to be no systematic preference for finding spatial matches with
SNRs in empty fields. However, we note that the search of the Green
catalog is heavily biased toward the innermost region of the Galactic
plane and lacks SNR detections at separations larger than a few
degrees from the plane.  A systematic study of existing surveys
probing large scale radio structures in combination with higher
sensitivity new single-dish and interferometric observations with
short baselines have the potential to shed more light on the nature of
radio emission found in those empty fields and whether it can be
related to any known class of objects that are candidates for
$\gamma$-ray production such as supernova remnants. This will be
investigated in a future publication. 

In addition, pulsars typically
show $\gamma$-ray photon indices below 1.8, however we only found two
empty fields with photon indices below 1.8. The average photon index
found is 2.34, which is consistent with the average spectral index of
flat spectrum radio quasars (FSRQ) and supernova remnants. However,
FSRQs are not expected to be very abundant at the probed $\gamma$-ray
luminosities, as can be inferred from Fig.~\ref{fig:vlbi}. The 83 identified 
pulsars listed in 2FGL have an average photon spectral index of 1.37$\pm$0.44, 
which is also incompatible with the distribution found for the empty fields. 
Thus, we suggest that a large fraction of the empty fields are either related
to emission from supernova remnants, possibly interacting with star
forming clouds in their vicinity, or a new unexpected Galactic
population of $\gamma$-ray emitters. The fraction of empty fields
among all unassociated $\gamma$-ray sources is indicated in
Fig.~\ref{fig:chart_end}.

In addition to the empty fields we found an abundance of sources
concentrated in the Galactic plane even among those $\gamma$-ray
sources for which we found radio sources within $3\sigma$ of their
localization but for which no compact AGN emission was detected. Both the 
distribution of empty fields and the location of new radio sources 
is biased by the selection effect that unassociated $\gamma$-ray
sources are concentrated within the inner $30^\circ$ of the Galactic
plane. However, we suggest that such a Galactic population could also
be related to emission from pulsars. \citet{2008ApJS..174..481L} have
shown that in dense regions of the Galactic plane, scatter broadened
pulsed radio emission can be detected in continuum observations up to
several GHz and could provide a significant contribution to the radio
sources that were found in this survey. We also consider scatter broadening 
as a possible reason for a lack of
non-detections of compact emission in the Galactic plane. In order to 
test this, we take the complete sample of known AGN detected by VLBI at
8 GHz above 180\,mJy as listed in the radio fundamental catalog (rfc-2014b) and compare the source
densities within $\pm 10^\circ$ of the Galactic plane. We found that within
the Galactic plane the source density is $5.9\cdot10^{-2}\,\mathrm{deg}^{-2}$ compared
to outside the Galactic plane where the source density is $7.0\cdot10^{-2}\,\mathrm{deg}^{-2}$.
Thus scatter broadening does not seem to play a significant role for VLBI 
detections and thus is not able to explain the lack of AGN $\gamma$-ray 
associations within the Galactic plane which supports the presence of 
a new Galactic population of $\gamma$-ray sources not yet associated.  A more 
thorough study of this effect is anticipated in future work focusing on the 
VLBI detections of all known $\gamma$-ray loud AGN.

We note that our search in the literature of reported associations and
a spatial cone search of the Green catalog showed that in the case of
2FGL\,J0553.9+3104 an association was made with PSR J0554+3107 and a
spatial match was found with SNR G179.0+2.6. In addition, we obtained
a high likelihood association ($\Lambda = 53.1$) with the AGN
J0553+3106.  In three other cases (2FGL\,J0538.1+2718, J0641.1+1006c,
and J2041.5+5003) where we found a spatial match with a SNR from the
Green catalog we also report a high confidence association with an
AGN. Thus, a spatial search among known supernova remnants produced a
possible 3\% false positive rate among the unassociated $\gamma$-ray
sources suggested to be related with supernova remnants. A more robust
method has to be developed in order to claim association with a
supernova remnant. We shoud note that the presence of an AGN
or a supernova remnant is not necessarily mutually exclusive and the
$\gamma$-ray emission is potentially a combination from both sources
in the respective fields. This shows the importance for a search of
AGN among $\gamma$-ray sources in the Galactic plane in order to
identify fields that have multiple $\gamma$-ray sources contributing
to the observed high energy emission.

Fujinaga, Y., Niinuma, K., Fijisawa K. et al. (2014, submitted) presented
at the European VLBI Symposium 2014 in October 2014 preliminary results 
of their independent program for observing 845 sources in the vicinity of 
231 2FGL unassociated sources. They have detected 28 sources with VLBI.
We confirm 19 of their detections.

\subsection{Comparison with indirect methods of association using continuum radio spectra}

The spectrum of synchrotron radiation from a source at parsec-scales
detected with VLBI is usually flat or inverted. The spectra at
kiloparsec scales have a significant contribution of the emission that
gives results from an interaction of a jet with interstellar or the intergalactic 
medium and thus are usually steep. Emission from a compact region dominates for
flat-spectrum sources and emission from extended regions dominates for
steep-spectrum objects, although there exists a considerable number of
exceptions in both cases. The $\gamma$-ray emission detected from AGN
is most likely related to the AGN core emission, due to a similar boosting 
of $\gamma$-rays and radio emission with common Doppler factors under small
observing angles.

The spectrum of a radio source is often cited as a criteria that helps
to associate it with a $\gamma$-ray source. We collected information
about the spectrum of radio sources using the CATS
database\footnote{\url{http://cats.sao.ru/doc/CATS\_list.html}}. That
database contains over 200 radio catalogs. Cross-matching these
catalogs, we computed spectral indices of
sources from observations at low resolution (10--$100''$). The results
are presented in Table~\ref{t:flat_spectrum}.  We should bear in mind
that these spectra are compiled from non-simultaneous data and thus
may be distorted by source variability.

\begin{table}[t!]
        \caption{Spectrum of \textit{Fermi} associations with AGN confirmed with VLBI.}
        \centering
        \begin{tabular}{lrl}
             \hline
             Flat  spectrum sources: & 830 (83\%) & $ \alpha > -0.5 \quad S \sim S^{\alpha}$ \\
             Steep spectrum sources: &  74 ( 7\%) & $ \alpha < -0.5 $ \\
             Unknown spectrum:       & 101 (10\%) & \\
             \hline
        \end{tabular}
  \label{t:flat_spectrum}
\end{table}

Indeed, $\sim$90\% of \textit{Fermi} associations with AGN confirmed
with VLBI have a spectral index flatter than $-$0.5. Among 76
associations found in the present paper 67 objects have a flat
spectrum, 7 have a steep spectrum, and for two objects the spectral
index has not been reliably determined. Of the 75 sources that were
not detected in our VLBI observations, 53\% were flat-spectrum
sources, 33\% were steep-spectrum sources, and others were without
known spectral index. If one wants to achieve completeness then
VLBI observations should be performed ignoring prior spectral information.

If a source has a flat radio spectrum, this does not automatically imply
it is an AGN or is associated with a $\gamma$-ray object. A number of 
planetary nebulae and HII compact regions show flat spectra as well.  
Since flat spectrum sources were predominantly observed in prior VLBI 
surveys, there is a certain selection bias that results in under-representation 
of steep spectrum sources. The spectrum of a source at low resolution 
does not reveal the nature of a source itself, but can serve as a proxy 
to make a rough estimate of the expected flux density from a compact component. 
However, the value of information about source spectrum is 
rather limited for population analysis since the ratio of flux density of a core to the 
flux density at kilo-parsec scales has a significant spread even for 
flat-spectrum sources. At the same time, this ratio is significantly higher 
than for the steep-spectrum sources. It is difficult to evaluate quantitatively
the probability of a \textit{Fermi} association using only estimates of
flux density at low resolution and a spectral index. However, in a
situation when observing resources are limited, reducing the list of
sources to be observed with VLBI by their pre-selection on basis of
the spectra, if such information is available, increases the number of
detections.

\subsection{Comparison with indirect methods of association using IR colors}

\citet{massaro_2011} cross-matched \textit{Fermi} sources associated
with AGN with the WISE catalog \citep{2010AJ....140.1868W} and found that matches occupy a
distinctive zone in the color-color diagram [3.4]--[4.6] ${\mu}m$
versus [4.6]--[12] ${\mu}m$.  This prompted them to suggest that this
feature can be used for association of \textit{Fermi} sources with
blazars. In numerous papers they claim their method provides an association 
with a probability of false association of just 3--4\%. In particular, in 
\citet{massaro_2013} they investigated two samples: NVSS and SUMSS sample of 324 unassociated
\textit{Fermi} sources (containing 56\% of unassociated 2FGL sources) and a
sample of 411 sources that we have observed in the first ATCA campaign. 
They proposed a number of associations.

\begin{table*}[htbp!]
      \caption{Performance of the different association methods. The column
      labeled ``predicted'' is that of the Massaro et al. method; the ``confirmed''
      number is the number of Massaro et al. predictions that we have confirmed with VLBI;
      the false associations column is the number of Massaro et al. predictions
      we can refute, because we found another counterpart with high likelihood 
      ratio; the VLBI number is the number of VLBI detections (i.e., 
      confirmed AGN) in the sample.}
      \centering
      \begin{tabular}{lrrrrr}
         \hline
                           & sample size & predicted & confirmed & false assoc. & VLBI \\
         NVSS+SUMMS sample & 324         &       52  &        30 & 2            & 95   \\
         ATCA\#1 sample    & 411         &       11  &         3 & 1            & 51   \\
         \hline
      \end{tabular}
  \label{t:massaro}
\end{table*}

In Table~\ref{t:massaro} we summarize and compare our statistics with that 
presented in \citet{massaro_2013}. We see that the Massaro et al. method, 
first, significantly under-predicts associations with AGN, second, a significant 
number of its predictions are not confirmed with VLBI. We should note, that to 
date, we have followed-up with VLBI only 1/2 of our targets, and the actual 
number of associations with \textit{Fermi} is expected to be higher. This 
discrepancy prompted us to perform our own analysis of the WISE color-color 
diagram using the ALLWISE catalog \citep{2011ApJ...731...53M}.
  
\begin{figure}[t]
  \centering
  \includegraphics[width=\columnwidth]{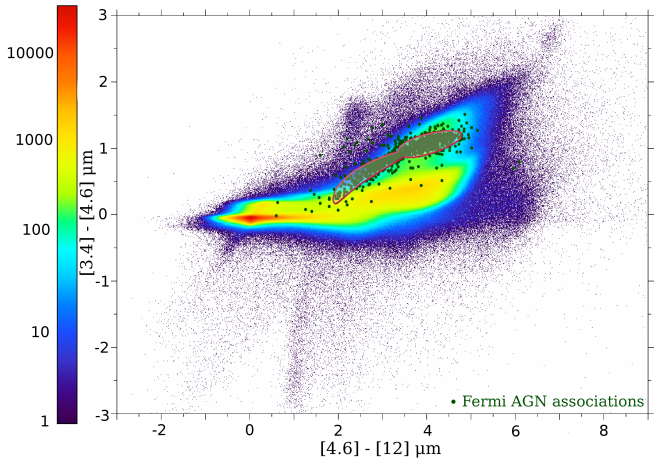}
  \caption{Color-color diagram of bright ALLWISE sources. Logarithm of source density
           is shown with color map. \textit{Fermi} sources associated with AGN with the 
            likelihood ratio $> 10$ are shown with green circles. The area of 
            \textit{Fermi} associations is shown with a red contour.
          }
  \label{f:wise}
\end{figure}

First, we selected a sample of \textit{Fermi} associations with AGN
confirmed with VLBI with the likelihood ratio greater than 10. In
total, 1005 sources fit this criteria. Among them, 959 objects have an
association with a source from the ALLWISE catalog that have been
detected at both 3.4, 4.6, and 12 $\mu$m within $5''$ of a VLBI
position. We find that \textit{Fermi} sources are associated with
predominantly bright ALLWISE objects: 97\% of the \textit{Fermi} sources
have a magnitude of less than 16.25 at 3.4~$\mu$m and 15.25 at
4.6~$\mu$m, well below the completeness level of ALLWISE. We discarded
fainter ALLWISE objects for further analysis. We build a color-color
diagram of differences in magnitude [3.4]--[4.6] $\mu$m versus
[3.4]--[12]~$\mu$m (Figure~\ref{f:wise}). The population of
\textit{Fermi} sources are indeed concentrated within a particular
region. We noticed that \textit{Fermi} sources concentrate along two ellipses 
in the modified diagram ([3.4]--[4.6]$)^2 \mu$m versus [3.4]--[12]~$\mu$m.
Therefore, we modeled the region with high concentration of \textit{Fermi} sources
with two 4th order figures that obey equations of 

$$
\Frac{(x-x_o)^2}{a^2} + \Frac{(y-y_o)^4}{a^4\, (1-e^2)^2} = 1
\label{e:el2}
$$
rotated counter-clockwise at angle $\zeta$. Parameters of the figures:
$ x_{o1} = 1.90, y_{o1} = 0.05, a_1 = 1.00, e_1 = 0.99, \zeta_1 = 27^\circ;
  x_{o2} = 3.42, y_{o2} = 0.88, a_2 = 0.78, e_2 = 0.92, \zeta_2 = 24^\circ$.
These figures capture 81\% of \textit{Fermi} sources with counterparts in the WISE catalog.

The area of \textit{Fermi} associations contains 777 out of 959 sources, 
i.e.\ 3/4 of the $\gamma$-ray objects that have associations with ALLWISE and 
have data at 3.4, 4.6 and 12~$\mu$m. The area contains 1.2 million ALLWISE bright objects
out of 68.5 millions. Only 0.06\% of them are known $\gamma$-ray
sources.  On average, 1.1 ALLWISE sources from the \textit{Fermi}
association area can be found in any field of radius $7.2'$ that is within
the upper limit $2\sigma$ error ellipse for 80\% of the \textit{Fermi}
sources, i.e. the probability to find at least one background source
is 67\%. This rules out using the color-color diagram for a
meaningful association without additional information.

Massaro et al. suggested using the association with the
lower frequency radio catalogs NVSS or SUMSS. Almost all known 
\textit{Fermi} associations with AGN are present in the NVSS or SUMSS 
catalogs.  Since the number of sources in radio catalogs is several
orders of magnitude fewer than there are in WISE, pre-selection of only
those WISE sources that have a counterpart in catalogs with frequencies
below 5\,GHz allows one to improve the predictive power of the WISE color-color
diagram. We selected NVSS sources brighter than 10~mJy at 1.4~GHz---
the NVSS catalog is complete at the level of 4~mJy --- and counted the
number of sources with bright ALLWISE counterparts and among them, the
number of sources that fall within the \textit{Fermi} association
area.  The results are summarized in Table~\ref{t:massaro1}. We note,
the statistics are different whether or not the Galactic plane is included.
Within $10^\circ$ of the Galactic plane 7.6\% of the sources have
colors that fall in the $\gamma$-ray AGN area, while the share of these sources
is a factor of 3~greater if we consider sources further than $10^\circ$ of the 
Galactic plane. We attribute this disparity to a
greater share of Galactic objects visible with both NVSS and ALLWISE at
low Galactic latitudes. Within the Galactic plane, 1.8\% of NVSS sources
with colors in the $\gamma$-ray AGN area account for 62\% of the
associations, and 4.2\% of the NVSS sources with $|b|>10^\circ$ with
colors in the same area account for 78\% of the associations.

\begin{table}[t]
   \caption{First two rows: statistics of NVSS sources brighter 
            10mJy at 1.4~GHz with $\delta > -40^\circ$ in the zone 
            of Galactic plane and in the zone beyond the Galactic plane. 
            Last two rows: statistics of \textit{Fermi} sources in these zones.}
   \par\medskip\par 
   \centering
   \begin{tabular}{lrr}
      \hline
                                         & $|b|>10^\circ$ & $|b|<10^\circ$ \\
      Total number                       & 478,808      & 102,288 \\ 
      \hline
      \# of ALLWISE counterparts            &  84,747      &  24,818 \\
      \# sources in the $\gamma$-ray AGN area        &  20,229      &   1,885 \vspace{2ex} \\
      Total \# of 2FGL sources           &     791      &      89 \\
      \# of 2FGL sources in the $\gamma$-ray AGN area &     615      &      56 \\
      \hline
   \end{tabular}
   \label{t:massaro1}
\end{table}

We performed a search for sources with colors in the $\gamma$-ray AGN
area for two samples: a) NVSS and SUMSS sources brighter than 10~mJy
at 1.4 and 0.8~GHz respectively within $2\sigma$ error ellipse of all
2FGL unidentified sources; and b) ATCA and VLA sources from our
program that are brighter than 10~mJy at any sub-band within
5--9~GHz. We restricted the flux density because weaker sources are not
reliably associated using our method. The results are shown in
Table~\ref{t:massaro2}.
  
\begin{table}[t]
   \caption{Statistics of two samples of radio sources found within $2\sigma$ error ellipses
            of \textit{Fermi} unidentified sources. Second column: sample of NVSS and SUMSS
            sources brighter than 10\,mJy at 1.4 and 0.8\,GHz respectively. 
            Third column: sample of sources brighter than 10\,mJy at 5--9\,GHz
            from our dedicated ATCA and VLA surveys.
            }
   \par\medskip\par 
   \centering
   \begin{tabular}{lrr}
      \hline
                                          & NVSS+SUMSS    & ATCA+VLA \\
      Total number                        &     1627     &  275     \\
      \hline
      \# of ALLWISE counterparts          &      431     &  160     \\
      \# sources in $\gamma$-ray AGN area &       96     &   64     \\
      \# VLBI observed sources            &       44     &   42     \\
      \# VLBI detected sources            &       43     &   41     \\
      \hline
   \end{tabular}
   \label{t:massaro2}
\end{table}

First, we should note that among our 95 VLBI detected sources, 76 are
confirmed associations. However among those sources only 41 fall
within the $\gamma$-ray area of the color-color diagram. Thus, the
method of identifying AGN counterparts of \textit{Fermi} sources using
their IR colors misses one half of the objects. At the same time, analysis
of the entire sample of confirmed radio-counterparts showed that only
3/4 of the sources have IR colors in the $\gamma$-ray AGN area. This
discrepancy can be explained by either an unintended bias in our VLBI
observation program or by a greater share of sources that are not
found in the $\gamma$-ray AGN area among radio weak sources, since the
sources that we have observed in our program are systematically
weaker. We see the detection rate of sources selected on the basis
of flux density at 5--9~GHz and IR colors is exceptionally high, more
than 95\%, unlike what would be expected.

It is worth mentioning that among 9,100 AGN detected in VLBI surveys only 30\% 
lie in the AGN area of the IR color-color diagram in contrast to 77\% of AGN 
that have detectable $\gamma$-ray emission. This implies that the color-color
diagram confines not the AGN population in general, but a specific sub-class 
of AGN. 

To summarize, the main differences between our method of association 
and the method of Massaro et al. are  1) we provide a quantitative measure for 
association confidence: the likelihood ratio; 2) the Massaro et al. color-color 
method misses half of the sources for which we find associations, however they
only focused on the population of $\gamma$-ray blazars and not all types of AGN
 as we do; 3) our method allows us to build a flux-limited sample of associations 
 suitable for population studies.

\section{Summary and Conclusions}\label{sec:5}

We present a catalog of radio sources resulting from an all-sky
radio survey between 5 and 9 GHz of fields surrounding the
localization of all unassociated $\gamma$-ray sources listed in the
\textit{Fermi} Large Area Telescope Second Source Catalog. In total we found
865 radio sources which could be considered as candidates for
association. Follow-up observations with VLBI on 170 selected from
those candidates led to the firm association of 76 previously unknown
$\gamma$-ray AGN. We provide the likelihood ratio for each
association.

In addition, among the 588 unassociated $\gamma$-ray sources targeted,
we found that in 129 not a single radio source was found above the
detection limit of our observations within the $3\sigma$ localization
error of the $\gamma$-ray sources. These ``empty'' fields were found to
be particularly concentrated around the innermost region of the
Galactic plane and we suggest them to be associated with a previously
unknown Galactic population of $\gamma$-ray emitters.

We compare our method of direct radio observations with indirect
methods of association based on IR colors as suggested by Massaro et
al. We have confirmed one half of their associations. We find that
other VLBI detected sources occupy a different place in the
IR color-color diagram and thus cannot be associated with AGN using only
their IR colors. This finding demonstrates the practical limit of the
indirect method.

Compared with other published results of $\gamma$-ray source association,
our approach appears to be the most productive for establishing AGN associations.
We plan to continue radio observations of unassociated sources that will be
discovered in subsequent releases of the \textit{Fermi} source catalog

\acknowledgments

We thank Dave McConnell for his support
with the ATCA observations. We thank members of the \textit{Fermi} collaboration, in particular
Francesco Massaro, Jean Ballet, and Elizabeth Ferrara, for useful discussions. We thank Dave Green,
Heinz Andernach, and the anonymous referee for useful discussions, corrections, and
comments that have significantly improved the quality of this publication.
We thank NASA for support under FERMI grant NNX12A075G. YYK was partly supported 
by the Russian Foundation for Basic Research (project 13-02-12103) and the Dynasty Foundation.  
The National Radio Astronomy Observatory is a facility of the National Science Foundation operated under cooperative
agreement by Associated Universities, Inc. The Australia Telescope
Compact Array / Long Baseline Array are part of the Australia
Telescope National Facility which is funded by the Commonwealth of
Australia for operation as a National Facility managed by CSIRO.  
This publication makes use of data products from the Wide-field Infrared Survey 
Explorer, which is a joint project of the University of California, Los Angeles, 
and the Jet Propulsion Laboratory/California Institute of Technology, and NEOWISE, 
which is a project of the Jet Propulsion Laboratory/California Institute of Technology. 
WISE and NEOWISE are funded by the National Aeronautics and Space Administration.
This work made use of the Swinburne University of Technology software
correlator, developed as part of the Australian Major National
Research Facilities Programme and operated under licence. This
research has made use of NASA's Astrophysics Data System and has made
use of the NASA/IPAC Extragalactic Database (NED) which is operated by
the Jet Propulsion Laboratory, California Institute of Technology,
under contract with the National Aeronautics and Space
Administration. This research has made use of data, software and/or
web tools obtained from NASA's High Energy Astrophysics Science
Archive Research Center (HEASARC), a service of Goddard Space Flight
Center and the Smithsonian Astrophysical Observatory, of the SIMBAD
database, operated at CDS, Strasbourg, France, and the TOPCAT software
version
4.1\footnote{\url{http://www.star.bris.ac.uk/$\sim$mbt/topcat/}}
\citep{2005ASPC..347...29T}. The authors made use of the database CATS 
\citep{2007HiA....14..636V} of the Special Astrophysical Observatory.

{\it Facilities:} \facility{VLA}, \facility{VLBA}, \facility{ATCA}, 
                  \facility{LBA}, \facility{FERMI (LAT)}.

\end{document}